\documentclass[fleqn,usenatbib]{mnras}

\usepackage{mathptmx}

\usepackage[T1]{fontenc}

\DeclareRobustCommand{\VAN}[3]{#2}
\newcommand{\noopsort}[2]{#2}
\let\VANthebibliography\thebibliography
\def\thebibliography{\DeclareRobustCommand{\VAN}[3]{##3}\VANthebibliography}

\usepackage{graphicx}	
\usepackage{amsmath}	
\usepackage{amssymb}	

\title[The predictive power of stochastic models]{The impact of stochastic
modeling on the predictive power of galaxy formation simulations}

\author[Josh Borrow et al.]{Josh Borrow,$^{1}$\thanks{E-mail: josh@joshborrow.com (JB)}
Matthieu Schaller,$^{2,3}$
Yannick M. Bah\'e,$^{3,4}$
Joop Schaye,$^{3}$
Aaron D. Ludlow,$^{5}$
\newauthor
Sylvia Ploeckinger,$^{2, 6}$
Folkert S.J. Nobels,$^{3}$
and Edoardo Altamura$^{7}$
\\
\\
$^{1}$Department of Physics, Kavli Institute for Astrophysics and Space Research, Massachusetts Institute of Technology, Cambridge, MA 02139, USA\\
$^{2}$Lorentz Institute for Theoretical Physics, Leiden University, PO Box 9506, NL-2300 RA Leiden, The Netherlands \\
$^{3}$Leiden Observatory, Leiden University, P.O.Box 9513, NL-2300 AA Leiden, The Netherlands \\
$^{4}$Institute of Physics, Laboratory of Astrophysics, Ecole Polytechnique F\'ed\'erale de Lausanne (EPFL), Observatoire de Sauverny, 1290 Versoix, Switzerland \\
$^{5}$International Centre for Radio Astronomy Research, University of Western Australia, 35 Stirling Highway, Crawley, Western Australia, 6009, Australia \\
$^{6}$Department of Astrophysics, University of Vienna, T\"urkenschanzstrasse 17, 1180 Vienna, Austria \\
$^{7}$Jodrell Bank Centre for Astrophysics, Department of Physics and Astronomy, The University of Manchester, Oxford Road, Manchester M13 9PL, UK \\
}

\date{Accepted XXX. Received YYY; in original form ZZZ}

\pubyear{2021}

\begin{document}
\label{firstpage}
\pagerange{\pageref{firstpage}--\pageref{lastpage}}
\maketitle

\begin{abstract}
All modern galaxy formation models employ stochastic elements in their sub-grid
prescriptions to discretise continuous equations across the time domain. In this
paper, we investigate how the stochastic nature of these models, notably star
formation, black hole accretion, and their associated feedback, that act on
small ($<$ kpc) scales, can back-react on macroscopic galaxy properties (e.g.
stellar mass and size) across long ($>$ Gyr) timescales.  We find that the
scatter in scaling relations predicted by the EAGLE model implemented in the
SWIFT code can be significantly impacted by random variability between
re-simulations of the same object, even when galaxies are resolved by tens of
thousands of particles. We then illustrate how re-simulations of the same object
can be used to better understand the underlying model, by showing how
correlations between galaxy stellar mass and black hole mass disappear at the
highest black hole masses ($M_{\rm BH} > 10^8$ M$_\odot$), indicating that the
feedback cycle may be interrupted by external processes. We find that although
properties that are collected cumulatively over many objects are relatively
robust against random variability (e.g. the median of a scaling relation), the
properties of individual galaxies (such as galaxy stellar mass) can vary by up
to 25\%, even far into the well-resolved regime, driven by bursty physics (black
hole feedback) and mergers between galaxies. We suggest that studies of
individual objects within cosmological simulations be treated with caution, and
that any studies aiming to closely investigate such objects must account for
random variability within their results.
\end{abstract}





\section{Introduction}

Cosmological galaxy formation simulations are one of the key tools in our chest
to understand the factors and processes that influence the evolution of galaxies
and their environments. By tracing the path of billions of resolution elements
at a simulation speed much faster than real time, these simulations allow for
detailed numerical experiments to be performed that are impossible through
observation alone due to the long timescales (typically $\gg$ Myr) involved in
galaxy formation. Modern simulations can track the evolution of hundreds of
thousands of galaxies simultaneously, allowing for self-consistent studies of
the interstellar, circumgalactic, and intergalactic media, over length scales
spanning five (or more) orders of magnitude from hundreds of megaparsecs down to
sub-kiloparsec scales, and have been successful in reproducing a huge swathe of  trends in
observed galaxies \citep{Dubois2014, Vogelsberger2014, Schaye2015,
Pillepich2018, Dave2019, Vogelsberger2020, Feldmann2022}. Despite this, the scatter
in scaling relations, for instance the ratio between the stellar mass and halo mass
of galaxies, remains poorly understood \citep{Matthee2017}.

Even with all of their successes, cosmological simulations still have limited
numerical resolution and as such many processes pivotal in the galaxy formation
process occur on scales smaller than those that are directly simulated. This has
led to the development of `sub-grid' models for a number of key processes
processes, notably: gas cooling \citep[atomic scales, e.g.][]{Wiersma2009,
Richings2014, Smith2017, Ploeckinger2020}, star formation \citep[sub-parsec
scales, e.g.][]{Springel2003, Schaye2008}, stellar feedback \citep[sub-parsec
scales, e.g.][]{DallaVecchia2012}, black hole accretion \citep[parsec scales,
e.g.][]{Bondi1952, Springel2005b, Rosas-Guevara2015, Tremmel2017,
Angles-Alcazar2017}, and black hole feedback \citep[parsec to kiloparsec scales,
e.g.][]{Booth2009, Weinberger2017}.

While the processes governing gas cooling (line emission, collisional
excitation, etc.) are relatively well-understood and can be accurately captured
across huge ranges in temperature, density, and metallicity through spectral
synthesis codes like Cloudy \citep{Ferland2017}, provided we assume chemical and
ionisation equilibrium, the underlying physics behind star and black hole
formation and feedback is significantly murkier \citep[for instance, we do not
yet have a fully predictive model for the initial mass distribution of
stars; see e.g.][]{Grudic2021}. This is further complicated by the limited spatial and
temporal resolution available in galaxy formation simulations, with coarse
graining of these processes required not only in space, but also in time.

Stochastic models are typically introduced for star formation, stellar feedback,
black hole growth, and black hole feedback. For a given scalar quantity $A$
(e.g. mass), it is typical to construct a growth rate $\dot{A}$ (e.g.  star
formation rate from a \citet{Schmidt1959} law, or black hole accretion rates
from a \citet{Bondi1952} prescription). It is then seemingly straightforward to
calculate the change in properties of the field over a discrete time-step
$\Delta t$, with $\Delta A = \dot{A} \cdot \Delta t$.  Now considering a
realistic case of a star formation law, and typical particle mass resolutions of
$M_{\rm g} \sim 10^6$ M$_\odot$, with time-steps of $10^{4-6}$ yr, this
implies that individual parcels of gas need to have star formation rates
$\dot{M}_* \sim 1-10^{2}$ M$_\odot$ yr$^{-1}$, comparable to that of an
entire $M_* \sim 10^{10}$ M$_\odot$ galaxy, for it to be consumed entirely
within a single time-step. This then leads to two possible solutions: either we
produce stellar resolution elements with masses much less than the gas particles
\citep[which may affect model calibration, increase memory consumption, lead to
ambiguities assigning particle softening lengths, or accelerate spurious energy
transfer between particles;][for example]{Ludlow2021, Wilkinson2022}, or we only
form stars stochastically. In the stochastic scenario, we re-write our growth
equations in the following form:
\begin{align}
    P(\Delta t) = \frac{\dot{A}}{A} \Delta t,
    \label{eqn:consumption}
\end{align}
where $P(\Delta t)$ is the probability that a discrete resource (usually an
entire gas particle) is consumed in the discrete time-step $\Delta t$. Then,
each resource draws a random number from 0 to 1, compares this against the
probability from equation (\ref{eqn:consumption}), and decides whether or not it
should be consumed. This means that, for instance, even gas particles with an
extremely high star formation rate may never produce a star if they are very
lucky (or unlucky, depending on your perspective). The disadvantage of this approach is then that there is a
potential offset in the timing of critical events in the history of a galaxy:
the birth of the first star, the first stellar feedback event, the first black hole
accretion and feedback events, along with all of their downstream impacts, which
will all be delayed relative to a continuous model.

Stochastic models then hence have a potentially significant dependence on the
specific choice of random numbers (i.e. the random seed employed), especially in
scenarios with poor resolution. We note that all choices of random numbers by
the stochastic models are valid. There is no `correct' timing of the initial
feedback events, for instance, in such a model. Most simulations use a single
realisation of these random models, with a single choice for random seed and a
single choice for the order of operations within the code. In this scenario,
there is a fixed single history for each halo and galaxy in the output, which is
an accurate and valid prediction from the model. Our goal is to understand the
implications and impacts of differing random number choices on the predictive
power of galaxy formation simulations - for instance, how uncertain are the star
formation histories of galaxies, given that our model only has predictive power
up to the random noise injected by the stochastic model?

Recent works by \citet{Keller2019} and \citet{Genel2019} have investigated the
impact of stochasticity on galaxy properties in isolated galaxy simulations and
full cosmological volumes, respectively. \citet{Keller2019} found that in
extremely isolated galaxies, the random variability in simple cumulative
properties of galaxies (notably the stellar mass) simply scales as a
Poisson-like error in that quantity (i.e. the random variability in the stellar
mass scales as $\sigma_{M_*} \propto M_*^{1/2}$). More complex scenarios, such
as mergers between galaxies, can significantly increase the variability in these
properties, by up to half a dex, though self-regulation of galaxies (for instance
through stellar feedback) can act as an attractor on long timescales and reduce
variability in galaxy properties. As the strength and level of self-regulation
varies between feedback models, so does their ability to control random
variability in galaxy properties, with stronger feedback models \citep[e.g. a
superbubble feedback implementation;][]{Keller2014} able to better control the
level of random variability between resimulations than weaker feedback
\citep[e.g. a blastwave feedback implementation;][]{Stinson2006}, as stronger
feedback models can often more tightly regulate star formation.

In contrast to most studies investigating `random' variability,
\citet{Genel2019} instead chose to run the {\tt Arepo} code with the
Illustris-TNG model in a (binary) reproducible mode\footnote{In this case, the
order of all operations on all quantities is consistent between re-simulations,
ensuring that each re-simulation should experience the exact same level of
round-off error.}, but pause the simulation at redshift $z=5$ to randomly perturb
the particle positions by an amount comparable to floating point round-off
errors.  These extremely small changes lead to dramatic changes in the
properties of individual galaxies, comparable to those found in
\citet{Keller2019}, with changes in the cumulative properties of galaxies
growing as a power law proportional to the square root of the evolved time over
the first Gyr of evolution. They also, notably, showed that galaxy properties at
$z=0$ were perturbed around the median in scaling relations (e.g. the
Tully-Fisher relation).

\citet{Davies2021, Davies2022} investigated stochasticity in the EAGLE
model \citep{Schaye2015} as a complicating factor for studies of so-called
``genetically modified'' galaxies \citep{Roth2016}. By using 9 re-simulations of
the same object (using a zoom-in technique) and only varying the random seed,
they found that instantaneous properties of galaxies, in particular the specific
star formation rate (sSFR), are significantly impacted by random variability
between re-runs of the same simulation. \citet{Davies2021} found that in
re-simulations, the same object can be classified as either star-forming (sSFR
> $10^{-11}$ yr$^{-1}$) or quiescent over a wide range of time (a span of
$\approx 6$ Gyr) in cases where the initial conditions are modified to promote
an early merger between galaxies. \citet{Davies2022} notes that gas-phase
properties, such as the mass fraction of the circumgalactic medium (CGM), can
vary by over an order of magnitude between random realisations of even the same,
unmodified, initial conditions when feedback from AGN is included. 


In this paper, we aim to quantify the impact that random variability between 
clones of the same simulation has on our ability to accurately model
galaxy scaling relations, as well as to investigate the origins of such 
variability in the EAGLE model. The rest of the paper is structured as
follows: in Section \ref{sec:swift}, we give an overview of the SWIFT
cosmological galaxy formation code and the SWIFT-EAGLE model used. Section
\ref{sec:matchhaloes} discusses our procedure for matching haloes across
`clone' simulations, and Section \ref{sec:scalingrelations} investigates
in detail the impact of random variability on the measurement of scaling
relations. Section \ref{sec:casestudy} considers the properties of a
single galaxy, matched across clone simulations, to further understand the
origin of variability in cumulative galaxy properties. In Section 
\ref{sec:galaxypropertyscatter} we further consider the increase in 
variability in galaxy stellar mass at high masses ($M_* >10^{10}$ M$_\odot$).
Finally in Section \ref{sec:conclusions} we summarise our main
conclusions.
\section{Simulation Setup}
\label{sec:swift}

\subsection{The SWIFT cosmological simulation code}

The simulations in this paper were performed with the SWIFT simulation code
\citep{Schaller2023}. SWIFT is a hybrid parallel code that utilises both thread
parallelism within a node and Message Passing Interface (MPI)-parallel
communications between nodes when required. In this paper, all simulations were
performed on a single high performance computing node, with 28 threads, and as
such no MPI-parallel component was used\footnote{The simulations
for this paper were performed with code revision {\tt v0.9.0-517-g75453d6f} on
the DiRAC COSMA7 system, on a single node with two Intel Xeon Gold 5120 CPUs.
The code was compiled with the Intel compiler version 18.0.20180210 with the
following CFLAGS: {\tt -idirafter /usr/include/linux -ip -ipo -O3 -ansi-alias
-xCORE-AVX512 -pthread -fno-builtin-malloc -fno-builtin-calloc
-fno-builtin-realloc -fno-builtin-free -w2 -Wunused-variable -Wshadow -Werror
-Wstrict-prototypes}}. To execute tasks simultaneously, SWIFT
leverages task-based parallelism, whereby individual tasks are placed in a
per-node queue, and then executed in parallel by threads that are assigned these
tasks - meaning different threads can be executing different pieces of physics
(e.g. hydrodynamics and gravity) simultaneously \citep{Schaller2016}.  This
differs from conventional galaxy formation simulation codes, that have typically
used branch-and-bound and data parallelism where every core executing the code
is simultaneously running the same code, and is designed to assist SWIFT in
producing excellent weak- and strong-scaling up to 10000 or more cores
\citep{Borrow2018}.

SWIFT includes multiple hydrodynamics and gravity schemes, but here we use the
SPHENIX SPH scheme, designed with galaxy formation sub-grid models in mind, for
the hydrodynamics \citep{Borrow2022}. For $N$-body gravity, we use a solver
employing the Fast Multiple Method \citep{Greengard1987} with an adaptive
opening angle, similar to \citet{Dehnen2014}.

\subsubsection{Random Numbers}

Given that random variations are the focus of this paper, we now describe in
detail how random numbers are generated within SWIFT.  Random number generators
for cosmological simulations should have high enough precision (be quantised
finely enough) to accurately model all processes, for instance the star
formation histories of galaxies, meaning that there must be a small enough
spacing between drawn random numbers to model extremely low probabilities. For a
process modelled every step in a simulation with a time-step of $\approx 10^4$
yr, across a Hubble time, a relative precision of at least $\approx 10^{-8}$ is
required, so that it only occurs on average once (e.g. a single gas particle
that must turn into a star).

SWIFT uses a random number generator based upon two POSIX functions for
generating random integers: {\tt rand\_r} and {\tt erand48}. Random numbers
produced by {\tt rand\_r} are 32-bit integers, making it unable to sample the
spacing between random numbers below $\approx 10^{-10}$, which is close to the
threshold required for even EAGLE-resolution cosmological simulations. Higher
resolution models will require even more precise random numbers, and as such
the 48-bit random number generator {\tt erand48} (spacing $\approx 10^{-15}$)
is used to supplement the typical {\tt rand\_r}.

Another (helpful, but not necessary) property of random number generators for
numerical workloads is that they should attempt to be reproducible based upon
local conditions. This means that the same random number should be generated for
the same particle at the same time-step, regardless of the number of time-steps
of the particle (or other particles in the volume) before this point. Consider a
single, shared, random number generator $R(n)$ where $n$ is the random
index of the random number that we wish to generate. This makes the random
number that a particle $i$ receives strongly dependent on all prior random
number generation events, and the order of particle processing within any given
step. One way to generate a pseudo-random, reproducible, number for a particle
is to make it dependent on the current time and particle index (i.e. one that
can be parametrised as $R(t, i)$).

In an effort to make the random numbers reproducible, the SWIFT random number
generator is effectively a hash containing four input values: the current unique
particle ID\footnote{In interactions, we generate a new unique ID given as the
combination of the two unique IDs ($A$ and $B$) of the two particles $AB + At +
Bt^2$ with $t$ the integer time. The integer time is a parameterization of the
internal simulation time within the code.}, a random number type identifier
(64-bit total), the current integer time (64-bit), and a 16-bit additional
random seed that we hold fixed here. This 144-bit buffer is then interpreted as
9 individual 16-bit unsigned integers ({\tt uint16}). We initialise a new {\tt
uint16} buffer with zero, and each of the 9 integers is XORed\footnote{Here by
XORed, we mean the exclusive or logic operation between two values $A$ and $B$
($A \nleftrightarrow B$), which is the same as the usual 'or' operation, just
with $A \nleftrightarrow B = 0$ if $A = B$.} with this buffer, and {\tt rand\_r}
is called to advance this seed. The new random number (which has been XORed nine
times, and progressed nine times through the random number generator), is used
to generate a new 144-bit buffer of 9 {\tt uint16}s.  The new 18 byte buffer is
then iterated in 3 individual 48-bit randomisations in a similar manner, now
using {\tt erand48} to generate the final 48-bit random seed. This final seed is
used with {\tt erand48} to generate a 48-bit mantissa from a uniform
distribution of floating point numbers in the range [0, 1).

\subsection{The SWIFT-EAGLE Model}
\label{sec:eaglexl}

SWIFT includes a fully open-source re-implementation of the equations solved in
the EAGLE model \citep{Schaye2015, Crain2015}, and a modified and modernised
version referred to as the SWIFT-EAGLE model\footnote{SWIFT is available for
download at \url{http://swiftsim.com}. The initial conditions and the required
scripts for performing the simulations in this paper are also available in this
same repository.}.

Throughout this paper, we will use the SWIFT-EAGLE galaxy formation model that
includes the components described below.

The SWIFT-EAGLE model includes the sub-grid radiative gas cooling and heating
prescription from \citet{Ploeckinger2020}\footnote{We use the file {\tt
UVB\_dust1\_CR1\_G1\_shield1.hdf5}.}. This model uses pre-calculated tables to
produce the equilibrium cooling and heating rates from the 11 most important
elements (H, He, C, N, O, Ne, Mg, Si, S, Ca, Fe) in the presence of a spatially
uniform and time varying UV background (UVB) based upon
\citet{Faucher-Giguere2020}.  The tables also include components from an
interstellar radiation field and cosmic rays and accounts for self-shielding.
More information on the prescription is available in \citet{Ploeckinger2020}. In
addition, the updates from \citet{PlanckCollaboration2020} were included in our
model with Hydrogen reionization occurring at a redshift of $z=7.5$.

As the simulation resolution used here is not high enough to track the evolution
of the cold and dense interstellar medium (ISM), we impose an effective pressure
floor (with $P \propto \rho^\gamma$) on gas following \citet{Schaye2008}. This
floor has gradient $\gamma = 4/3$ and is imposed at densities $n_{\rm H} >
10^{-4}$ cm$^{-3}$, normalised to $T = 8000$ K at density $n_{\rm H} = 0.1$
cm$^{-3}$.

Star formation is implemented following the \citet{Schaye2008} pressure law, in
a similar fashion to the original EAGLE model. Instead of the metallicity
dependent star forming threshold from \citet{Schaye2004}, SWIFT-EAGLE uses the
tabulated properties from \citet{Ploeckinger2020} to limit star formation to
cold gas. We assume that the (unresolved) cold gas phase is in pressure
equilibrium with the effective pressure that describes the ISM in SWIFT-EAGLE
and close to thermal equilibrium, i.e. the temperatures, where the net cooling
rates from the \citep{Ploeckinger2020} tables are zero. The density and
temperature pair that matches these conditions are referred to as sub-grid
properties ($T_{\rm sub-grid}$, $n_{\rm H, sub-grid}$). Gas particles with equal
pressure can have different sub-grid properties based on their species abundances
and therefore different thermal equilibrium functions. This adds an effective
metallicity dependence in the star formation threshold without specifying it
explicitly. Here, star formation is allowed for gas particles with $T_{\rm
sub-grid} < 1000$ K, or $n_{\rm H, sub-grid} > 10$ cm$^{-3}$ and $T_{\rm sub-grid}
< 10^{4.5}$ K. The latter condition is only relevant for extremely low
metallicity where the thermal equilibrium temperature is high in the
\citet{Ploeckinger2020} tables due to the added interstellar radiation field.
As discussed in the introduction, the star formation prescription in SWIFT-EAGLE
is stochastic, with each gas particle computing a probability of forming stars
based upon the pressure law, and the entire gas particle being converted to a
single star particle representing a simple stellar population if chosen.

Stellar feedback is again implemented stochastically following the prescription
in \citet{DallaVecchia2012}, with stars in the stellar population modelled with
a \citet{Chabrier2003} initial mass function with a mass range $0.1 < M_* / {\rm
M}_\odot < 100$. We assume that stars with a mass of $8 < M_* / {\rm M}_\odot <
100$ explode as core-collapse supernovae, releasing the typical $10^{51}$ ergs
of energy. We include the same feedback scaling function as EAGLE, as described
in \citet{Crain2015}, with free parameters for the metallicity and density
scaling, as well as minimal and maximal energy injection and efficiency
parameters (see Section \ref{sec:calibration}).

Supernova energy is injected thermally into the closest gas particle to the star
particle at injection time, following \citet{Chaikin2022}, and now include a
distribution in delay times between the birth of star particles and their
injection of supernova energy to sample the stellar lifetimes, a notable change
between the new model and EAGLE (which used a fixed delay time for all
supernovae). The coupling efficiency of stellar feedback to the ISM is modulated
through a number of free parameters, discussed in Section \ref{sec:calibration}.
Finally, following \citet{Wiersma2009b} and \citet{Schaye2015}, we include the
influence of type Ia supernovae and AGB stars (and their associated winds) on
their environment through energetic feedback, mass flux, and metal injection. We
allow gas particles to be split in two once their mass exceeds four times the
initial mean gas particle mass, as in galaxies with low gas fractions the mass
flux from these processes becomes significant.

Black hole formation and active galactic nucleus (AGN) feedback are implemented
following \citet{Booth2009} and \citet{Bahe2022}, with black holes initially
seeded within friends-of-friends (FoF) groups\footnote{The implementation of FoF
within SWIFT is discussed in \citet{Willis2020}.}. This seeding occurs in haloes
above a minimal mass $M_{\rm FoF} = 10^{10}$ M$_\odot$, and seed black holes are
given an initial sub-grid mass of $M_{\rm BH} = 10^4$ M$_\odot$. Black hole
seeding is a deterministic process; as soon as a halo crosses the minimal FoF
mass, and does not already host a black hole, it is seeded with one. Black holes
are additionally repositioned frequently towards the centre of potential within
their host galaxy, as described in detail in \citet{Bahe2022}.

Black holes grow by accreting mass from their surrounding gas particles and
through mergers with other black holes \citep[with our merger strategy described
in detail in][]{Bahe2022}. We employ the `nibbling' strategy as described by
\citet{Bahe2022}, where in most cases small fractions of the mass of nearby gas
particles are accreted. The accretion rates of black holes are goverened by the
\citet{Bondi1952} law, as described below.

AGN feedback is implemented following \citet{Booth2009}, where energy for
feedback is stored in a reservoir until a nearby gas particle can be heated by 
$\approx 10^{8.5}$ K. The coupling efficiency and heating temperature of AGN
feedback is given as a free parameter in the model, as described below. We note
here that we no longer employ the \citet{Rosas-Guevara2015} model to
suppress BH accretion rates depending on the angular momentum of the ambient
gas, as described in \citet{Bahe2022}.

\subsection{Model Calibration}
\label{sec:calibration}

The SWIFT-EAGLE model contains free parameters that must be calibrated to data.
For the original EAGLE simulations, this procedure was outlined in
\citet{Crain2015}, where crucially a density- and metallicity-dependent scaling
of the supernova efficiency were introduced to simultaneously fit the galaxy
stellar mass function and active galaxy mass-size relation. This calibration
procedure was performed by hand: changes in the model and free parameters that
were deemed to be physically reasonable were used to bring the model predictions
closer to fixed observations.

Now that a good baseline model parametrisation has been established and
re-implemented in the SWIFT code, it must be recalibrated to offset differences
between the two codes (hydrodynamics model, minimum stellar mass for
core-collapse supernovae, feedback injection strategy, cooling tables, black
hole accretion model, and softening changes following \citet{Ludlow2020}).
Simply re-running the original parameters from EAGLE leads to, among other
differences, an underestimation of galaxy sizes at around $10^8 < M_{*} / {\rm
M}_\odot < 10^{10}$.

The free parameters in the model considered here are as follows:
\begin{itemize}
    \item $f_{\rm E, min}$, the minimal feedback energy fraction.
    \item $f_{\rm E, max}$, the maximal feedback energy fraction.
    \item $n_{\rm H, 0}$, the density pivot point around which the feedback
          energy fraction plane rotates.
    \item $\sigma_{\rm Z}$ and $\sigma_{\rm n}$, the width of the feedback
          energy fraction sigmoid in metallicity and density dimensions.
    \item $\epsilon_{\rm f}$, the coupling coefficient of radiative efficiency
          of AGN feedback to the surrounding gas.
    \item $\Delta T_{\rm AGN}$, the AGN heating temperature.
    \item $\alpha$, a constant suppression factor for black hole accretion\footnote{
    We note that in many simulations $\alpha$ is used as an enhancement factor in AGN
    feedback, but in all cases here we choose $\alpha < 1$.}.
\end{itemize}
These free parameters enter into the following key equations of the model:
\begin{equation}
    f_{\rm E} = f_{\rm E, max} - \frac{f_{\rm E, max} - f_{\rm E, min}}{
        1 + \exp\left(-\frac{\log_{10} Z / Z_{\rm 0}}{\sigma_Z}\right) \cdot
        \exp\left(\frac{\log_{10} n_{\rm H} / n_{\rm H, 0}}{\sigma_{\rm n}}\right)
    },
    \label{eqn:energyfrac}
\end{equation}
which sets the energy injected by each supernova explosion as a multiplicative
factor to the fiducial $10^{51}$ ergs, moving between $f_{\rm E, min}$ and
$f_{\rm E, max}$ dependent on the density and metallicity of the surrounding gas
($n_{\rm H}$ and $Z$ respectively) that are candidates for being heated. We note
that in the original EAGLE simulations the birth density of the star was used in
this equation, rather than the density of the heated gas.

The growth rate of each black hole is given by the Eddington-limited Bondi
accretion rate,
\begin{equation}
    \dot{m}_{\rm BH} = \min\left[
        \alpha \frac{4 \pi G^2 m_{\rm BH}^2 \rho_{\rm gas}}{
            (c_{\rm s}^2 + v_{\rm gas}^2)^{3/2}
        },
        \frac{4\pi G m_p m_{\rm BH}}{\epsilon_{\rm r} c \sigma_{\rm T}}
    \right],
\end{equation}
where $G$ is Newton's constant, $m_{\rm BH}$ the mass of the black hole,
$\rho_g$ the ambient gas density, $c_{\rm s}$ the sound speed as measured around
the black hole, $v_{\rm gas}$ the bulk velocity of the gas relative to the black
hole, $m_{\rm p}$ the proton mass, $c$ the speed of light, $\epsilon_{\rm
r}=0.1$ the radiative efficiency of feedback, and $\sigma_{\rm T}$ the Thomson
cross-section.

The feedback energy associated with each black hole accretion event depends upon
the accreted mass in each time step, $\Delta m$, as:
\begin{equation}
    \Delta E = \epsilon_{\rm r} \epsilon_{\rm f} \Delta m c^2.
\end{equation}
This energy is stored in a reservoir carried by each black hole until it is
possible to heat the nearest gas particle by the energy corresponding to a
temperature increase of $\Delta T_{\rm AGN}$.

The free parameters in the SWIFT-EAGLE sub-grid model were calibrated using
emulators employing the Gaussian Process Regression-based python module 
SWIFTEmulator \citep{Kugel2022}. This process uses re-simulations of the same
EAGLE-L00025N0376 volume (henceforth referred to as EAGLE-25) also used here,
with parameters drawn randomly to fill a Latin hypercube \citep{Schaye2015,
Crain2015}.  More information on the specifics of the emulation procedure will
be described in Borrow et al. (\emph{in prep}), and so here we simply give a
brief outline and reasoning for the choice of free parameters.

The free parameters were all calibrated simultaneously through four successive
waves of emulation, and the following values were found to be the best available
fit to the galaxy stellar mass function and galaxy mass-size relation when
considering our 25 Mpc volume:
\begin{itemize}
    \item $f_{\rm E, min}=0.388$,
    \item $f_{\rm E, max}=7.37$,
    \item $n_{\rm H, 0}=0.412$ cm$^{-3}$,
    \item $\sigma_{\rm Z}=0.311$,
    \item $\sigma_{\rm n}=0.428$,
    \item $\epsilon_{\rm f}=0.035$,
    \item $\Delta T_{\rm AGN}=10^{8.62}$ K,
    \item $\alpha=0.645$.
\end{itemize}
Significantly more information on this calibration process, and the model
impacts, will be available in Borrow et al. (\emph{in prep}), though in the
current work we do demonstrate the performance of these parameters on several
key galaxy scaling relations.

\subsection{Simulation Volumes}

To investigate the impact of run-to-run variations on the predicted properties
of galaxies, 16 identical `clone' simulations were performed. These simulations
used the well-studied `EAGLE-25' initial conditions \citep{Crain2015}, which
evolves a 25 Mpc$^3$ simulation volume initially containing $376^3$ dark matter
particles ($m_{\rm DM} = 9.77\times10^6$ M$_\odot$) and initially $376^3$ gas
particles ($m_{\rm g} = 1.81 \times 10^6$ M$_\odot$), starting at redshift
$z=127$. The same cosmology as in the original EAGLE simulation from Planck-13
\citep{PlanckCollaboration2014} was used, to maintain same initial conditions as
prior work. This is a spatially flat Lambda-CDM cosmology with dimensionless Hubble
parameter $h=0.6777$, cosmological constant density parameter $\Omega_\Lambda =
0.639$, baryon density parameter $\Omega_{\rm b} = 0.048$, clustering amplitude
$\sigma_8=0.8288$ and spectral index $n_{\rm s} = 0.9611$.

\begin{figure}
    \centering
    \includegraphics{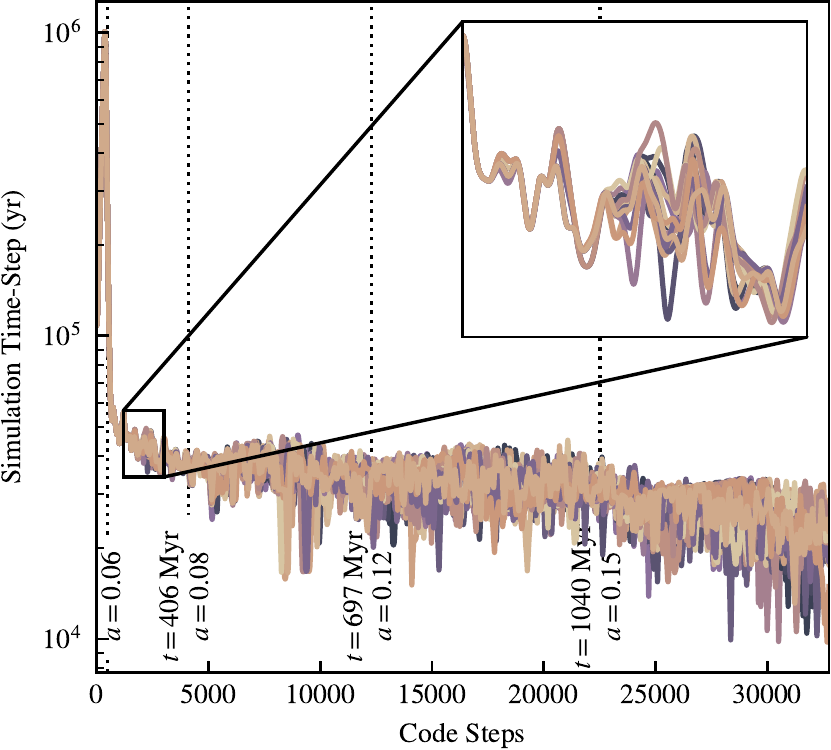}
    \caption{Evolution of the global time-step of the simulation over the course
    of the simulation (with each line a different clone simulation),
    demonstrating that even with a fixed random seed the non-deterministic
    parallelism of SWIFT causes divergences in the time integration as early as
    a few hundred Myrs into the simulation. Time-steps have been smoothed using
    a gaussian kernel over the nearest 32 to highlight the long-term trend. Once
    the time-step has deviated between the simulations, they will draw different
    random numbers (and there is no reasonable way to keep such number
    synchronised), leading to uncontrollable deviation in the impacts of star
    formation and feedback. Timing lines are only representative as they show
    the current time at that step number for a random clone simulation, apart
    from the first line at $a=0.06$ which represents the time ($\approx 300$
    Myr) at which the first star particle is formed. We additionally note that
    all of the lines presented here are valid and accurate. Simulations that
    only consider one realisation of their model would simply choose one of
    these lines (or another equally valid set of time-steps).}
    \label{fig:scalefactorupdates}
\end{figure}

These volumes were evolved to redshift $z=0$, with full particle snapshot dumps
at redshifts $z=$ 7, 5, 4, 3, 2, 1.5, 1, 0.75, 0.25, 0.2, 0.1, 0.05, 0.01,
0.001 and $0.0$. These snapshots were then analysed with the VELOCIraptor halo
finder \citep{Elahi2019}, producing group catalogues based on the 3D Friends of
Friends (FoF) algorithm, that were then further processed using the
swiftsimio toolchain \citep{Borrow2020, Borrow2021a}.

\subsection{Impact of Non-Determinism}

Many factors can impact the (non-)deterministic nature of a simulation code,
some of which are relatively hidden to users. The use of non-blocking and
asynchronous communication patterns (as in SWIFT) may lead to different arrival
times of data, dependent on network conditions, that could then lead to changes
in results. Even synchronous, blocking, communications may have significant
non-determinism if they are implemented in a non-deterministic way within the
chosen communication library, or the number of communication nodes is changed.
In practice a single-threaded code that has an entirely deterministic set of
algorithms can have its result depend on infrastructure choices such as the
instruction set available on the processor, compiler, or even compilation
options. Single Input Multiple Data (SIMD) instruction sets (such as the AVX2
and AVX-512 instruction sets available on our machine) and aggressive
compilation options (such as {\tt -O3} or {\tt -ffast-math}), can (for instance)
change the order of operations and unroll loops for a performance benefit. These
changes lead to different assembler output and hence different round-off errors
for the exact same input code, even when using the same compiler, compared to an
un-optimised counterpart. As such, we employ the exact same compiled binary in
all of our tests (including the dark matter-only simulation, which in SWIFT can
be performed without the need to re-compile).

Due to the non-deterministic nature of SWIFT, it was not necessary to change the
random seeds in the clone simulations in order to investigate the impact of
stochasticity. Fig. \ref{fig:scalefactorupdates} shows the variation in the
current time step, the time interval over which the active set of particles are
being evolved, as a function of the number of steps.  After only a few thousand
steps (at $z \approx 15$, before significant sub-grid physics effects begin) the
evolution of the simulations diverge due to differing round-off errors within
the calculation as expected, and due to the feedback from the first stars. This
is despite the fact that all time-steps across all clones are discretised in a
similar manner, using the typical power-of-two hierarchy \citep[see
e.g.][]{Hernquist1989, Borrow2021b}.

As the random number generator takes as input the current step, from the point
at which the various simulations diverge in their step-simulation time-step
curves, each clone will naturally use a distinct sequence of random
numbers\footnote{If we insert a new time-step $t_{\rm new}$ by splitting $\Delta
t_2$ in half for particle $i$ into a previously established sequence $\{t_1,
t_2, \dots, t_n\}$, then our random number sequence will become $\{R(t_1, i),
R(t_{\rm new}, i), R(t_2, i), \dots, R_(t_n, i)\}$. Though the majority of these
random numbers will be binary identical, because the \emph{sequence} has changed
the evolution of particle properties for particle $i$ will diverge.}. This
effect can spread throughout the volume due to the time-step limiter employed in
the hydrodynamics, which does not allow neighbouring particles time-steps to
differ by more than a factor of four \citep{Borrow2022}. We stress here that
there is no `correct' series of time-steps that the code should take to evolve
to the `true' answer. Each simulation takes a valid path that is dependent on
small variations in input properties into the time-step calculation that diverge
due to differing round-off errors in the calculation. All simulations follow a
similar average path in this figure, with the time step decreasing with the
number of steps taken as denser structures build up within the volume.

\begin{figure}
    \centering
    \includegraphics{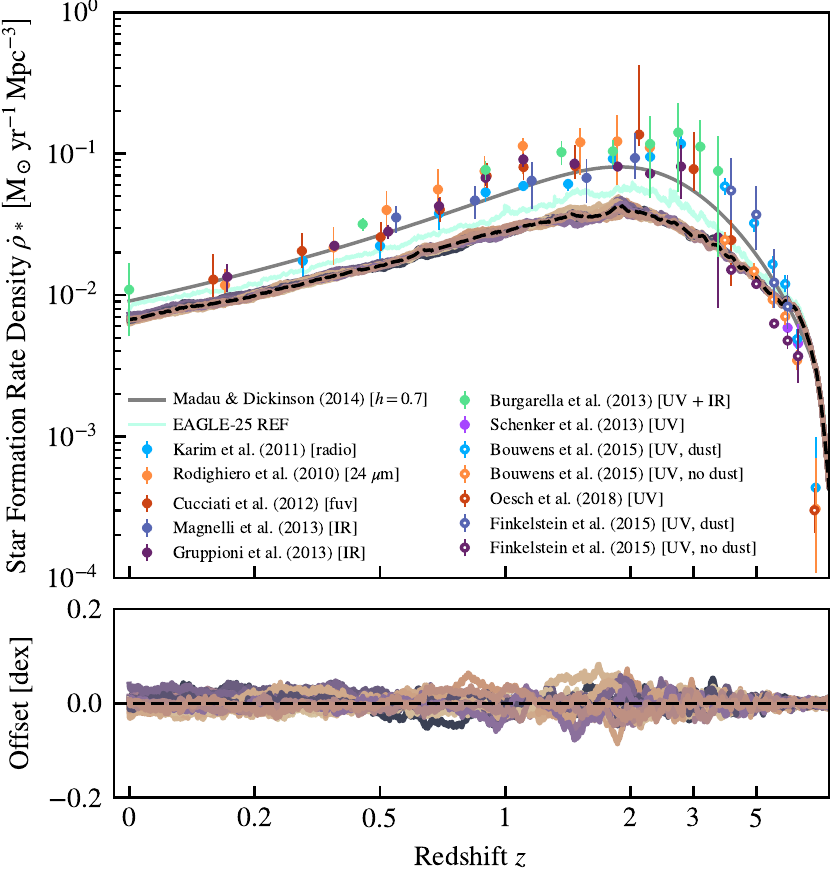}
    \caption{The global (cosmic) star formation rate density history for the
    various clone simulations (top) and deviations from the median (bottom).
    Each line shows a different clone, with the black dashed line showing the
    median across all clones. There is very little (less than 0.1 dex) scatter
    in this relation across all time, with the variation peaking when the cosmic
    star formation rate peaks.  Data in the background comes from multiple
    sources, including \citet{Madau2014, Karim2011, Rodighiero2010,
    Cucciati2012, Magnelli2013, Gruppioni2013, Burgarella2013, Schenker2013,
    Bouwens2015, Oesch2018, Finkelstein2015}. In addition, we show the same
    volume (EAGLE 25 REF; aquamarine line) simulated with the original EAGLE
    code and reference model as in \citet{Schaye2015}.}
    \label{fig:globalpropertiessfr}
\end{figure}

Fig. \ref{fig:globalpropertiessfr} shows the variation in the global (cosmic)
star formation rate density (SFRD) as a function of cosmic redshift, with the
same line colours as Fig. \ref{fig:scalefactorupdates}. We additionally show the
median across the clones at each redshift as the black dashed line, and the
offset of each clone from this median in the bottom panel. For comparison, we
show the original EAGLE Ref model \citep{Schaye2015, Crain2015, Furlong2015} on
the same EAGLE-25 volume, at the same resolution, in aquamarine. The systematic
offset between our clone simulations and this original EAGLE model
\citep[simulated with a modified version of the Gadget code;][]{Springel2005} is
due to the recalibration of model parameters when moving to SWIFT (see \S
\ref{sec:eaglexl}), as well as the aforementioned model changes. Both
simulations undershoot the observational data shown at $z < 3$.

Fig. \ref{fig:globalpropertiessfr} shows that variations between the clone
simulations are small and typically less than 0.1 dex. The scatter in the SFRD
peaks when the star formation rate peaks, at around $z=2$. The level of scatter
in this relation is strongly dependent on the simulation volume size;
considering an infinitely large volume, variations between the growth of
individual haloes will be washed out in this globally averaged metric.

\begin{figure}
    \centering
    \includegraphics{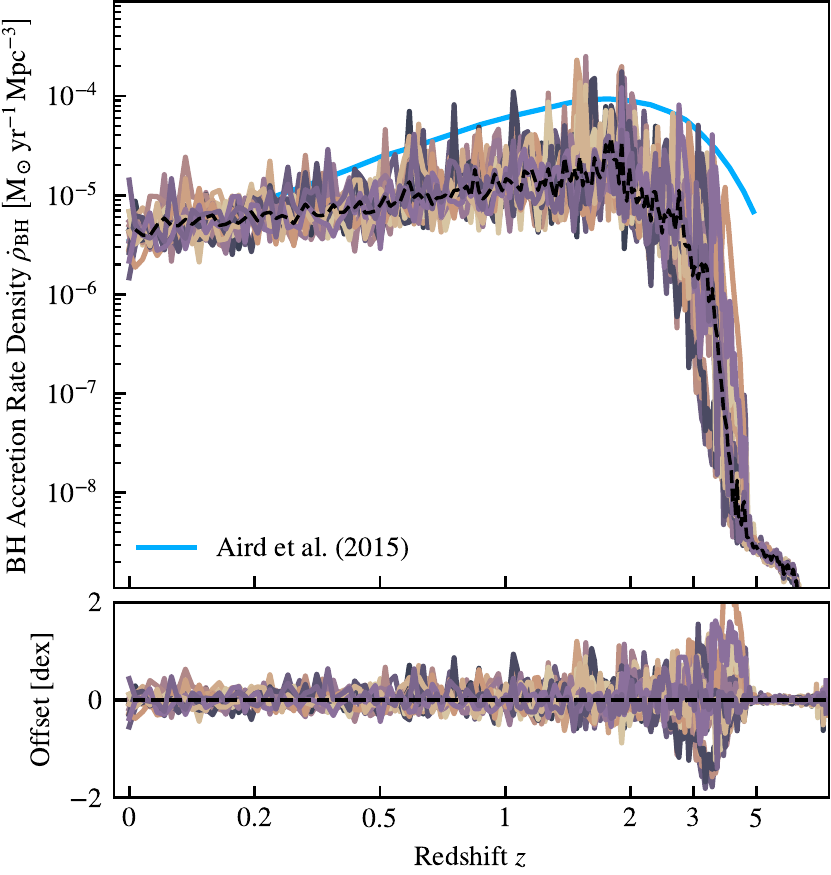}
    \caption{The global black hole accretion rate density (BHARD) history for
    all clone simulations (top) and variation from the median (bottom).  Each
    line shows a different clone, with the black dashed line showing the running
    median. There is significant variation between clones, shown in the lower
    panel, but this is consistent with the high level of variability
    of the BHARD on short timescales within an individual simulation. In the
    background, we show observational data from \citet{Aird2015}, spanning the
    range $z=0.0-5.0$.}
    \label{fig:globalpropertiesbhard}
\end{figure}

Fig. \ref{fig:globalpropertiesbhard} shows the variation in the
\emph{instantaneous} global black hole accretion rate density (BHARD) between
the clone simulations. The variability in this quantity is significantly higher
($\approx$ 2 dex) than for the global star formation rate density. This can
primarily be attributed to two main (connected) reasons: the much poorer
sampling of black holes (at $z=0$ there are of order thousands, compared to the
millions of star particles that have effectively sampled the SFRD), and the high
cadence variability of the BHARD within an individual clone simulation
\citep[see also][for further work on this in the EAGLE model]{McAlpine2017,
Nobels2022}. Even between these 16 clones, we are not able to extract a smooth
and stable median BHARD, suggesting an extreme level of variability.


In summary, there is significant variability between
the clone simulations, but volume averages can wash this out even in small boxes.
\section{Matching Haloes}
\label{sec:matchhaloes}

To investigate how the properties of individual galaxies are affected by random
differences between simulations, haloes must be matched between the clones.
Matching haloes in a context where the aim is to find differences between each
halo that is matched presents a problem; how can we develop a method that allows
for robust matching, producing few false positive matches, and allows for
potentially substantial differences in the mass content of haloes between
clones?

There are a number of strategies for matching haloes between simulations,
usually employed to match dark matter only simulations to baryonic counterparts.
A common strategy is to use the $n$ most bound particles of each halo, and find
the halo in another simulation within which they reside
\citep[e.g.][]{Velliscig2014, Schaller2015, Bose2018, Bose2019, Lovell2018,
Genel2019}. This strategy can have issues when trying to match haloes that are
undergoing a merger, which we will see are crucial to understanding the impact
of stochasticity. Instead, we employ a strategy that matches haloes by their
co-moving final-state position in the volume.

To match haloes between clone simulations, we first match all haloes in each
clone with a single, common, dark matter-only `parent' simulation. We choose to
do this as stochastic effects in dark matter-only simulations have been shown to
be significantly smaller than those in full hydrodynamical simulations
\citep{Genel2019}. Then matches in the clone simulations are back-propagated
through matches with the same halo in the dark matter-only parent. Matching via a
dark matter-only parent also prevents one of the clone simulations being
considered the `main' simulation and allows for fair comparisons between all
clones. Only central haloes (i.e. not satellites) are matched between
simulations, and within the rest of the paper unless explicitly stated all halo
properties correspond to those of the central subhalo.

\begin{figure}
    \centering
    \includegraphics{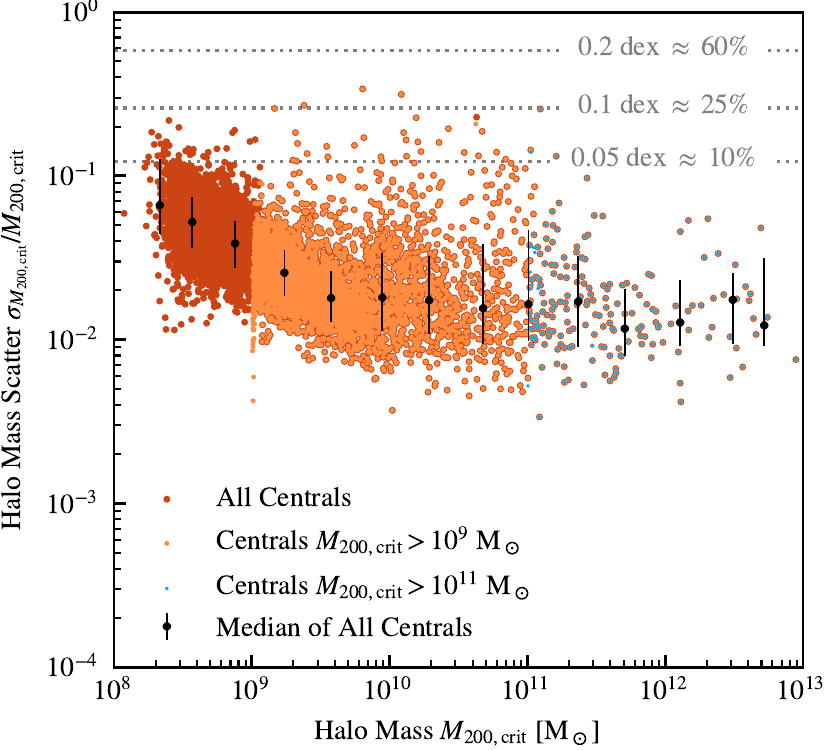}
    \caption{Reduced scatter in halo mass $M_{\rm 200, crit}$, defined as the
    ratio between half of the 16-84 percentile range of masses ($\sigma_{M_{\rm
    200, crit}}$), for three different mass cuts in our matching procedure. Red
    points show matches for all central haloes, orange for haloes with $M_{\rm
    200, crit} > 10^9$ M$_\odot$, and blue for haloes $M_{\rm 200, crit} >
    10^{11}$ M$_\odot$.  Matches show an extremely high level of consistency
    between these mass cuts.  Black points show the median and 16-84 percentile
    range of halo mass scatters. Horizontal dashed lines show, as a guide,
    representative scatter as a percentage of the original halo mass and the
    corresponding log-space (dex) offsets.}
    \label{fig:combinedspatialhalomassscatter}
\end{figure}

Haloes are matched using their comoving positions in the volume. The centres of
the haloes in the dark matter only simulation, as defined by their most bound
particle, are used to build a KDTree. This tree is then searched using the
positions of all haloes in each clone simulation in turn for the nearest 10
haloes. Then, the halo that matches the mass ($M_{\rm 200, crit}$\footnote{Here
we define $M_{\rm 200, crit}$ as the mass within a volume $4\pi R_{\rm 200,
crit}^3 / 3$ with a density of 200 times the critical density $\rho_{\rm crit}$
at that epoch, centered on the most bound particle within each halo.}) of the
parent halo within 50\% and has a distance between the two haloes of less than
the $R_{\rm 200, crit}$ of the parent is considered to be the match. For haloes
to be included in any analysis they must be matched to at least half of the
clone simulations (which in practice removes only a handful of haloes). As we
will see later, our strategy naturally leads to no conflicts, but when there are
some we choose the halo that is closest in mass. In the following discussion
we aim to demonstrate how robust our matching procedure is, though one possible
drawback of this method is that it assumes there is a one-to-one matching between
central haloes\footnote{We have also performed the analysis with a typical
bijective particle matching method \citep[as in e.g.][]{Sawala2013}, and
although we find a higher abundance of low-mass matches, the results for the
highest masses (M$_* > 10^9$ M$_\odot$) are almost identical.}.

Fig. \ref{fig:combinedspatialhalomassscatter} shows how the scatter in halo
mass, defined as half of the 16-84$^{\rm th}$ percentile range of a quantity
matched across clones, correlates with the median halo mass matched across
clones. Throughout the paper we use this metric, written as
\begin{equation}
    \sigma_{x} = \frac{{x}_{{\rm P}_{84}} - {x}_{{\rm P}_{16}}}{2},
\end{equation}
where ${x}_{\rm{P}_{y}}$ is the y$^{\rm th}$ percentile of the distribution of
variable $x$, to quantify the scatter between clone simulations. In most cases,
we show the reduced quantity (i.e. $\sigma_{x} / x$) where we reduce by the
median value matched across clones. We vary the minimal mass at which we match
haloes (increasing the distance between haloes), to demonstrate the accuracy
of our matching methodology.

At the lowest halo masses, scatter is significant, up to 10\% of the total mass
of the halo (and more in extreme cases). This is likely due small differences in
the coordinates of individual particles that place them just outside of the
$R_{\rm 200, crit}$ aperture, as at the lowest mass ($3\times10^{8}$ M$_\odot$),
only around 30 particles make up the halo. As the halo mass increases, this
`shot noise' decreases to a fixed level of around 2\%.

The first thing to note from this figure is that there are no points close to
the cut-off mass ratio (50\%), meaning that there is not a significant set of
haloes being rejected from the matching procedure simply because they have a
high intrinsic scatter in halo mass between clones (and hence relative to the
dark matter only parent).

\begin{figure}
    \centering
    \includegraphics{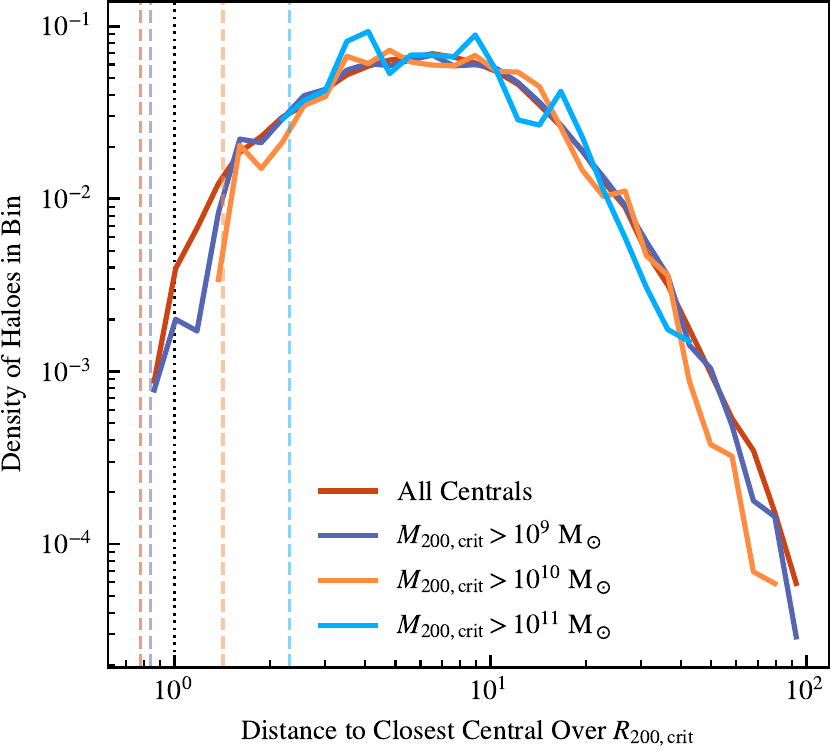}
    \caption{Distribution of distances to the nearest central halo for all halo
    centres (in units of $R_{\rm 200, crit}$) in the dark matter only simulation
    above different halo mass thresholds (coloured lines).  Dashed vertical
    lines show for each cut the minimum distance in the entire volume. Even when
    resolved haloes of any mass are considered (red line), only a fraction of a
    percentage point of haloes have a separation less than $R_{\rm 200, crit}$.
    This indicates that if this snapshot was matched with itself a cut at $R /
    R_{\rm 200, crit} = 1$ alone would prevent almost all false positive matches
    where we incidentally match a halo with one containing different particles.
    Hence, false positive match detections require significant movement of
    haloes (i.e. further than $R_{\rm 200, crit}$) between the dark matter only
    parent and the clone simulations.}
    \label{fig:haloseparationratiohistogramfull}
\end{figure}

The second thing to note is the almost perfect consistency between matching as
the mass threshold is increased. Increasing the mass threshold naturally
increases the minimum distance between field haloes and reduces the number of
false positives.  This ensures confidence that our matching procedure is robust,
even if the properties of the galaxies resident in these haloes show significant
variation in their properties.  For all parent haloes the closest separation
between field haloes is almost always larger than the match distance criterion
(see Fig.  \ref{fig:haloseparationratiohistogramfull}), even across the mass
range.

\begin{figure}
    \centering
    \includegraphics{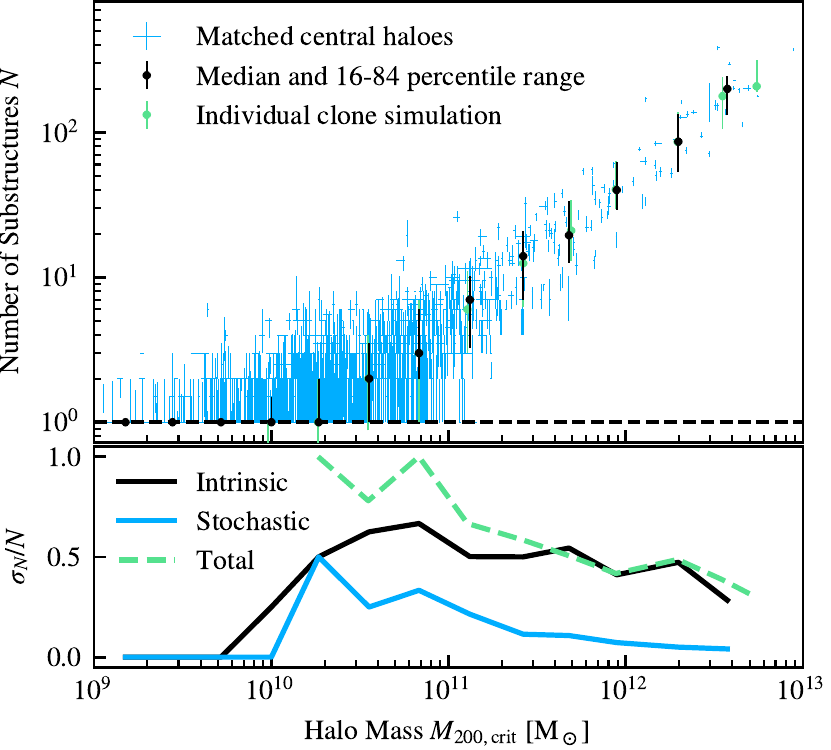}
    \caption{\emph{Top panel}: the number of substructures for each central as a
    function of host halo mass. Each blue cross represents the scatter amongst
    clones (16-84 percentile range), with the cross being centered on the
    respective medians.  The black points and error bars show the median of
    these medians in 14 equally log-spaced bins, and the length of the bars is
    the 16-84 percentile scatter of the medians within the bin. The black dashed
    line shows the extreme $N=1$ case, where the central is the only
    substructure. The green points show the scaling relation recreated from an
    individual clone simulation. \emph{Bottom panel}: the scatter, coloured
    analogously, from the top panel. The blue line shows the mean (binned)
    scatter in individual haloes (i.e. the mean length of the blue vertical
    lines), the green dashed line shows the scatter in the scaling relation as
    recreated in an individual clone (i.e. the length of the green error bars),
    and the black line shows the scatter in the medians of the matches (i.e. the
    length of the black error bars).}
    \label{fig:num_substructure}
\end{figure}

Fig. \ref{fig:num_substructure} shows the number of resolved substructures within each
halo as a function of halo mass. Haloes contain more substructure as the mass
increases, as expected \citep[e.g.][]{Kravtsov2004}, though this quantity is
clearly resolution-dependent, as substructures can only be tracked down to a
minimum of 30 particles, or $M_{\rm 200, crit} \approx 3\times10^8$ M$_\odot$.

In this case we see both large levels of scatter between clones (i.e. the
scatter for a given halo, shown by the blue lines, is large) at low mass, and
large intrinsic scatter (i.e. the scatter in the median, shown by the black
lines) at high mass. The number of substructures considered bound to a given
halo can easily vary by 100\% at the low-mass end ($M_{\rm 200, crit} \approx
10^{10}$ M$_\odot$), with this meaning that the halo has either a single
additional substructure or not. This can occur if the merger time is different
between clones, or if a very small substructure is just above or below the
minimum mass for the halo finder.

At the high-mass end ($M_{\rm 200, crit}\approx 10^{12}$ M$_\odot$), the scatter
in this relation is mainly dominated by differences between the individual
haloes' accretion histories and substructure abundances. The scatter between
clones is very small (less than 5\%), whereas there can be nearly half a dex of
variation within each mass bin. As such, within this regime, we would expect the
accretion and merger history of individual bound haloes to be similar between the
clone simulations.
\section{Galaxy Scaling Relations}
\label{sec:scalingrelations}

\begin{figure*}
    \centering
    \includegraphics{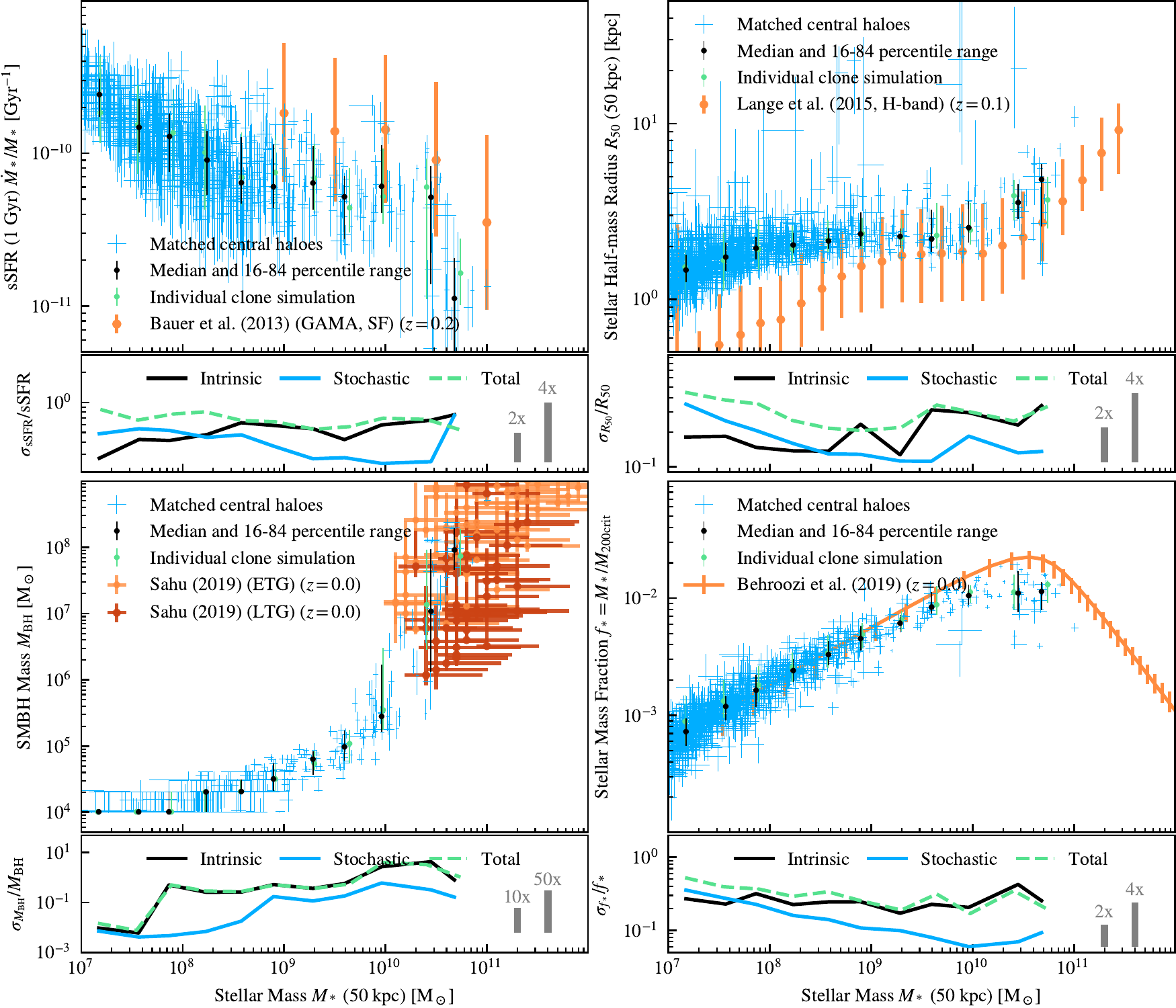}
    \caption{Four key scaling relations, all at $z=0$ and as a function of
    galaxy stellar mass: specific star formation rate (sSFR; top left, computed
    from the star particles and averaged over 1 Gyr), galaxy size (stellar
    half-mass radius within a fixed 50 kpc aperture; top right), black hole mass
    (most massive black hole; bottom left), and the halo stellar mass fraction
    ($f_*$; bottom right), plotted, for reference, with various observational
    datasets. Blue crosses show the run-to-run scatter, representing the 16-84
    percentile range for each matched halo (crossing at the median across
    clones). Black points show medians and 16-84 percentile ranges in bins of
    stellar mass for the medians of the individual halo matches. Green points
    show the scaling relation binned in a similar fashion, but for only one
    randomly selected clone.  The smaller, lower, panels in each subplot each
    show three lines representing the scatter in the medians of the matches
    (i.e.  representing the same size as the error bars in the black points, an
    estimate of the intrinsic scatter) in black; in blue, the median individual
    scatter in each bin (i.e.  the median size of the blue error bars, our
    estimate for the scatter due to stochastic effects); and as green dashes the
    total scatter in an individual scaling relation measured from a single
    clone.}
    \label{fig:scalingrelations}
\end{figure*}

In Fig. \ref{fig:scalingrelations} four key galaxy scaling relations are shown,
with each blue cross representing the 16-84 percentile scatter amongst the clone
simulations for matched haloes, with the cross placed at the median point. In
black, we show the binned scaling relation (of the medians across clones) and
its 16-84 percentile scatter as an indication of intrinsic halo-to-halo scatter
when `random' variations are smoothed over.  Finally, in the background we show
a selection of observational data as a visual reference. As these
data are at various redshifts, not matched with the simulations, we show them
primarily to demonstrate differences between the scatter in the observed
relations relative to the random variability of individual haloes. Each of the
properties is shown as a function of galaxy stellar mass, calculated within a
fixed 50 kpc aperture.

The smaller panels in Fig. \ref{fig:scalingrelations} show how the scatter in
three cases changes with galaxy stellar mass. In blue, we show the median value
(within 14 equally log-spaced bins across the stellar mass range) of the scatter
between clone simulations, to demonstrate the typical random variability of
galaxies due to stochasticity. This represents half of the median vertical
length of the blue crosses in the top panels. In black, we show the scatter in
the intrinsic scaling relation, representing the scatter in the average
properties (across clones) of the galaxies. This corresponds to half of the
length of the vertical black lines in the upper panel.  In a green dashed line,
we show the scatter of the scaling relation measured in a randomly chosen clone
simulation, which arises from both intrinsic and stochastic scatter. This can be
thought of as, on average, adding in quadrature the offset from the median line
to the center of the thin blue crosses (the median across clones for that
object) with a displacement from this median, bounded roughly by the extent of
the thin blue lines.

The top left panel of Fig. \ref{fig:scalingrelations} shows the specific star
formation rate (sSFR$= \dot{M}_* / M_*$) of galaxies against their stellar mass.
The star formation rates were recalculated from the birth times of stars in each
galaxy to reconstruct a 1 Gyr averaged star formation rate. This was done to
smooth over short-timescale fluctuations in the instantaneous star formation
rates. 1 Gyr also happens to be of the order of the halo dynamical times, and as
such the star formation rate on these timescales should mainly be driven by the
accretion history, and should be relatively insulated against small-scale
variations that are susceptible to individual round-off differences
\citep{Iyer2020}. We compare our sSFR relation against the \citet{Bauer2013}
results from the GAMA survey, which only includes galaxies that are considered
to be star forming, (sSFR$>10^{-11}$ yr$^{-1}$). Here, we include all central
galaxies to investigate the quenching behavior at stellar masses $M_* >
10^{10}$ M$_\odot$.

The first notable property in the sSFR relation is that when using a standard
quenching definition (of sSFR$<10^{-11}$ Gyr$^{-1}$) some of the clone galaxies,
with fundamentally similar accretion histories (note the typical 2\% variation
in halo mass from Fig. \ref{fig:combinedspatialhalomassscatter} and less than
5\% variation in substructure abundance at high mass in Fig.
\ref{fig:num_substructure}), may be considered quenched and others active.

The vertical scatter in the sSFR of individual galaxies can reach up to 1 dex,
though the average scatter of clone galaxies (blue line in the lower panel) is
significantly lower than the intrinsic scatter in the scaling relation (black
line). Throughout the mass range shown, the scatter in the measured scaling
relation in an individual clone is dominated by the intrinsic scatter (black
line), except at the lowest masses of $M_* < 10^9$ M$_\odot$. In this regime,
the two components of scatter combine, leading to significantly larger measured
scatter than the intrinsic level. The other exception is at the very highest masses,
where the stochastic variability rises rapidly due to the onset of variable quenching.

The top right panel shows the stellar mass-size relation, compared against data
from the GAMA survey \citep{Lange2015}. Here, we use the projected stellar
half-mass size of galaxies measured within a fixed 50 kpc aperture. Sizes of the
lowest-mass galaxies ($M_* \lessapprox 10^9$ M$_\odot$) are artificially
increased in the simulation due to spurious size growth from sampling noise in
gravitational interactions between stars and dark matter \citep{Ludlow2019,
Wilkinson2022}.  Within this regime we also see increased scatter between the
clone galaxies, with the scatter between clones dominating over the intrinsic
scatter. It is foreseable that the processes leading to inflated galaxy sizes
(spurious gravitational heating, oversoftening of sizes), especially in a regime
where there are relatively few ($< 1000$) particles resolving each galaxy, could
influence the scatter between clones.

There are a number of galaxies significantly above the median, with extremely
large (close to 1 dex) scatter in their sizes. These galaxies are typically
ongoing mergers, with the central in some cases being identified alone (i.e.
post-merger, with all galaxies collected into one central), and in other cases
the two merger components being identified separately. We will come back to this
point in \S \ref{sec:casestudy}. At the highest masses, we see that the scatter
in galaxy size returns to a low level comparable to the scatter in stellar mass,
with the overall scatter in each scaling relation corresponding to the intrinsic
scatter.

The bottom left panel shows the galaxy stellar mass-black hole mass relation.
There is significant scatter in this quantity between the clone simulations for
black holes in the rapid growth phase \citep[$10^6 < M_{\rm BH} / {\rm M}_\odot
< 10^8$ and $10^9 < M_* / {\rm M}_\odot < 10^{11}$][]{Bower2017, McAlpine2018}
of up to 1 dex. This does not correspond with significant scatter in
stellar mass, though, with the largest scatter in stellar mass appearing at the
high-mass end, as confirmed in the bottom right panel showing the stellar to halo
mass ratio.

In almost all cases, the scatter in $M_{\rm BH}$ is dominated by the intrinsic
scatter in the scaling relation, but this is misleading in the range $10^{10} <
M_* / {\rm M}_\odot < 10^{11}$ where the gradient of the scaling relation is
large within the single bin leading to a large predicted scaling relation
scatter. The average scatter in black hole mass in this regime is approximately
equal to the difference between the current and prior median at one dex lower
stellar mass. Finally, the average clone-to-clone scatter reduces at the
high-mass end rather sharply and shows significantly less variation than the
comparison scatter data from \citet{Sahu2019}, though measurements of black hole
mass are highly uncertain with errors of greater than 1 dex in many cases.

In the bottom right panel, we show the relationship between the stellar mass of
galaxies and their halo stellar mass fraction, defined as the ratio between the
galaxy stellar mass and $M_{\rm 200, crit}$. We compare against results from the
2019 release of UNIVERSEMACHINE \citep{Behroozi2019}. In almost all cases, aside
from the very lowest masses, we see that the scatter in the stellar mass-halo
mass relation is dominated by the intrinsic scatter on average. However, we have
a number of galaxies that show up to 0.5 dex scatter in stellar mass even at
high masses ($M_* > 10^{10}$ M$_\odot$), which is in stark contrast to the
results from \citet{Keller2019} who showed that scatter in stellar mass should
decrease with increasing mass when considering isolated galaxies. The obvious
difference here is that we are not considering purely isolated cases, and
include AGN feedback. A number of high-scatter cases can originate from ongoing
mergers \citep{Davies2021}. In addition to mergers, we see that the average
clone-to-clone scatter in stellar mass fraction increases at the very highest
masses. It is unlikely that all of these galaxies are currently undergoing a
major merger, and we will revisit this point in \S
\ref{sec:galaxypropertyscatter}.

Our levels of scatter in stellar mass are roughly comparable to the results
presented in \citet{Genel2019}, who also employed full cosmological volumes
but used a different code and model, though at a similar resolution.
\citet{Genel2019} report that their average standard deviation (a measure
comparable to our scatter) between clone galaxies is of order 0.1 dex at $z=0$
for stellar mass and stellar half-mass size, though they designed their
simulations to include a perturbation at $z=5$, with the difference between
clones growing over time, with our clone-to-clone divergence beginning at
$z=15$.  They additionally see larger scatter in specific star formation rate
(also 1 Gyr averaged), typically around 0.2 dex, which is consistent with our
results. The similarity between the scatter in both (completely independent)
galaxy formation models, implemented in completely different codes, with
different parallelisation strategies, is remarkable.

As a general point, if the intrinsic scatter dominates, this leads to similar
scatter measured in the scaling relation from a single clone.  In cases where
the stochastic noise dominates, however, the total scatter is a combination of
both the intrinsic and noise scatter.
\section{A Case Study}

\label{sec:casestudy}

In this section the properties of a halo that has been selected to be
``uninteresting'' (with typical scatter in stellar mass) are explored to
investigate the origins of the scatter between clone simulations.

\begin{figure}
    \centering
    \includegraphics{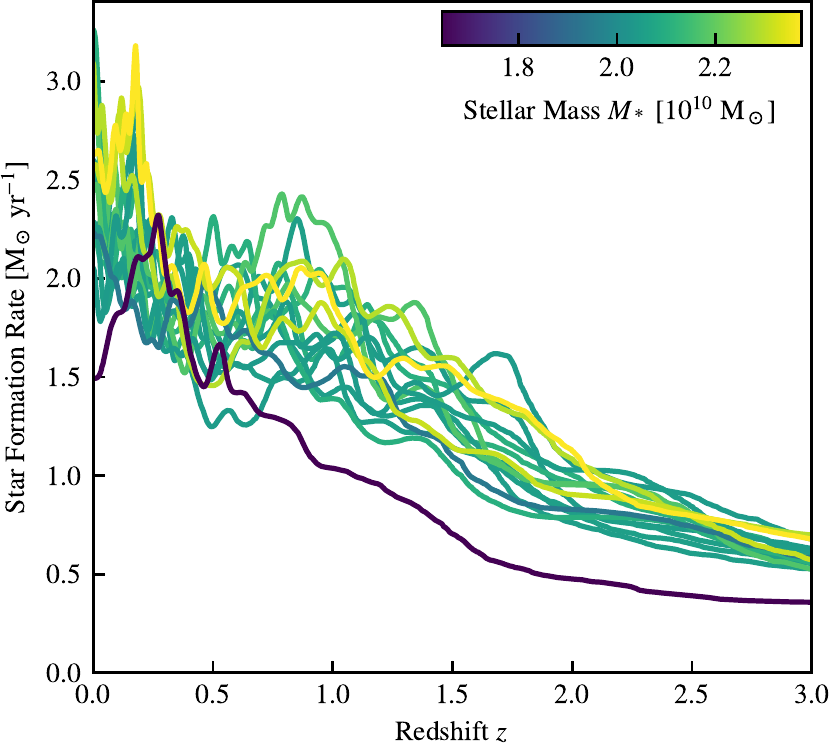}
    \caption{The star formation rate (averaged over the birth times of 256
    particles at each point, and smoothed using a Gaussian filter over the
    nearest 128 points for clarity) recovered from $z=0$ stellar particles, for
    a single galaxy with $M_* \approx 2 \times 10^{10}$ M$_\odot$ matched in
    each of the clone simulations. Each line is coloured by the $z=0$ stellar
    mass of the galaxy, which shows a variation of $\sim 10\%$. The outlier
    shown in dark purple is indeed matched correctly, and will be discussed
    further in this section.}
    \label{fig:differentialmassgrowth}
\end{figure}

In Fig. \ref{fig:differentialmassgrowth} we show the star formation rate of the
galaxy in each of the 16 clone simulations from $z=3$ to $z=0$. We recover the
star formation rate from the birth times of the star particles belonging to the
galaxy at $z=0$, calculating a moving average based upon the birth times of 256
particles to ensure each point is captured with the same Poisson error.

There is $\approx 0.1$ dex scatter in the final-state stellar mass (of $\approx
2 \times 10^{10}$ M$_\odot$) that is created through the varying star formation
rate across cosmic time.  The majority of the curves overlap significantly, with
$\approx 30\%$ variation in star formation rate at any given time. It is notable
that the highest mass galaxy (yellow line) does not have a continuously higher
star formation rate across time, but it has an above-average star formation rate
for the longest.

There is, however, an outlier (dark blue curve)
that has a significantly lower star formation rate (by about 25\%) across
much of the history of the galaxy, corresponding to a lower final stellar mass.
Additionally, it lacks the peaks between $z=2$ and $z=0$ shown in the other
curves, suggesting a suppressed merger history, which we will discuss in detail
in the following text.

\begin{figure*}
    \centering
    \includegraphics{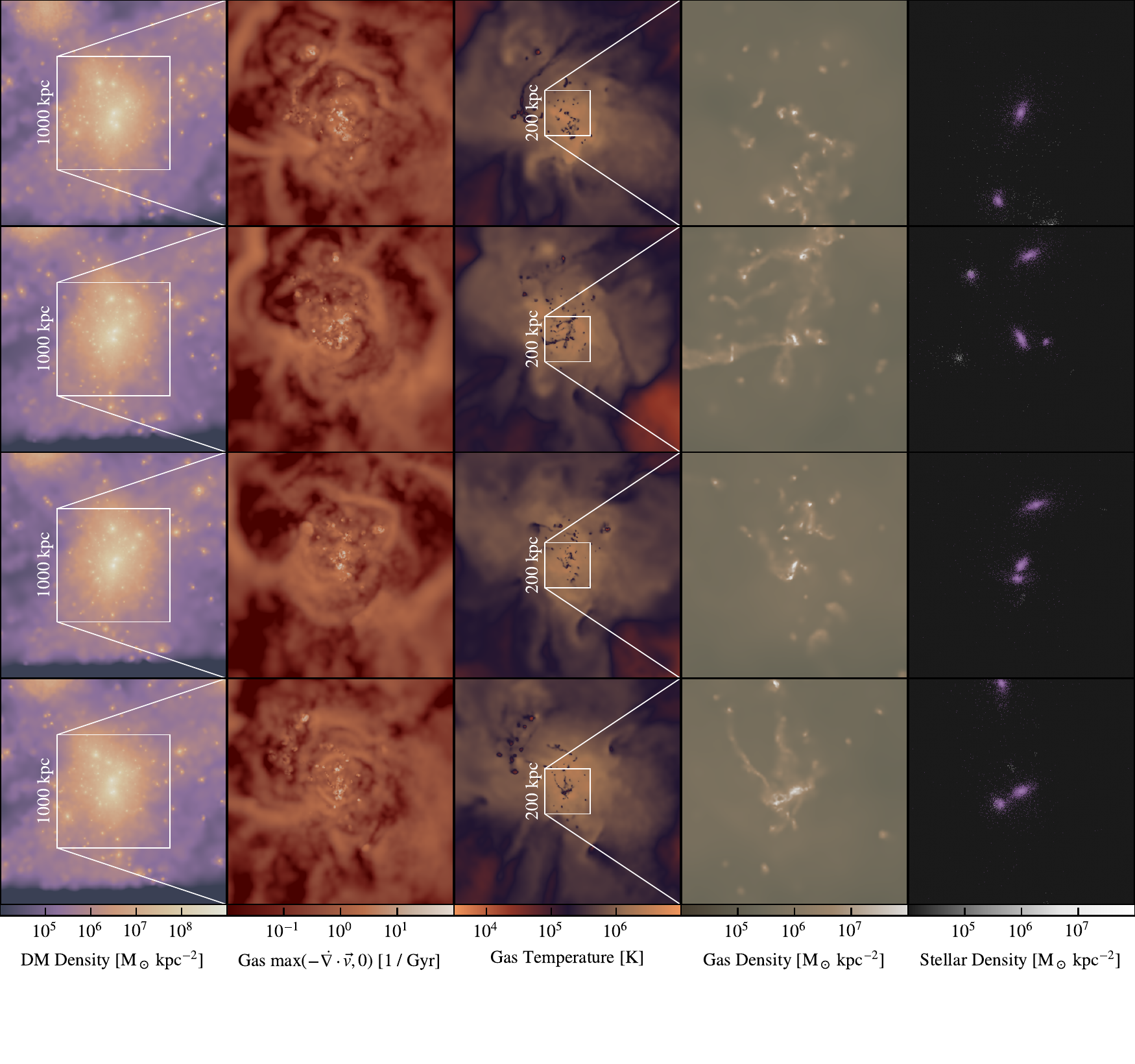}
    \vspace{-0.56in}
    \caption{Various properties of the case study halo at $z=0$. From left to
    right, we show the projected dark matter surface density (projection depth
    500 kpc), the convergent gas velocity divergence (a tracer of shocks,
    projection depth 100 kpc), the mass-weighted gas temperature (100 kpc), the
    gas surface density (100 kpc), and finally the stellar surface density (100
    kpc), zooming in from left to right. In the stellar density image, particles
    associated with the matched central halo are highlighted in pink, whilst
    particles not associated with the central halo are white. Clone haloes are
    ordered from lowest to highest mass from top to bottom.}
    \label{fig:casestudy}
\end{figure*}

In Fig. \ref{fig:casestudy} we display maps of different properties of four
clones of this galaxy as projections through the volume (projection depths are
indicated in the caption). From top to bottom we show a selection of the lowest
to highest mass clone galaxies whose star formation rates we shown in Fig.
\ref{fig:differentialmassgrowth}. Horizontally, the panels show the dark matter
projected density along the line of sight, the convergent gas velocity
dispersion (which highlights shocks), the projected gas temperature, and finally
the projected gas and stellar densities.

First, the dark matter density image (shown with a side length of 2 Mpc)
demonstrates that there may be different positions of substructures between
clones. In \citet{Genel2019}, it is shown that there is on average a difference
of $\approx 2$ kpc between dark matter particles at $z=0$ between clone
simulations, indicating that there is a significant impact on small-scale
dynamics that bleeds through to infall times of substructures as seen here (e.g.
the placement of the two substructures just inside the inset square at 11 and 12
o'clock). 

Zooming in to a 1 Mpc box side length (second column), we now turn to look at
the large scale gas dynamics. The value of $\max(-\dot{\nabla} \cdot
\vec{v}_{\rm g}, 0)$ is shown projected along the line of sight to highlight
shocks within the circumgalactic medium (CGM) and intergalactic medium (IGM)
around this galaxy. Though the general picture is similar in all four cases,
there are significant differences in the specific placement of shocks,
indicating differences in the large-scale gas dynamics and feedback timing
between clones (e.g. the bubble at 2 o'clock in the third row).

On the same scale we show the projected gas temperature along the line of sight
(third column), further indicating significant differences between the gas
dynamics in the CGM and IGM.  Though the size of the hot gas halo surrounding
the galaxy is well constrained across clones, the detailed structure, including
the direction of outflows, and the abundance and placement of cold gas clumps,
shows differences. In the top row, there is a substructure undergoing stripping
of its cold gas (9-12 o'clock, just outside the 200 kpc inset box), leaving a
trail within the CGM. The trail is completely absent in the other panels,
as the gas has broken up into hotter and more diffuse clumps.

The two rightmost columns show zoom ins of the structure of the galaxy disks
themselves, with an image of side length of 200 kpc. Here we begin to see large
differences in the merger timing between the clones, illuminating why the  star
formation rates of the galaxies are so different in Fig.
\ref{fig:differentialmassgrowth}. This final-state galaxy is an ongoing merger
between three individual bound objects at $z=0$. Not only are the galaxies in
the top row (lowest mass) smaller in size, they have proceeded less than their
clones into the merger process, with one of the three infalling galaxies (bottom
right, edge of the panel) not being included in the bound structure and hence
not being included as part of the star formation rate calculation (stars that
are included as `bound' to the central are highlighted in pink in the right
column, with this being essentially all of them except the incoming galaxy that
is mostly hidden in the top row, and the small dwarf at 8 o'clock in the second
row).

The ISM and CGM around these galaxies show significant differences between
clones. The morphology of dense inflows seems to be completely unconstrained
between clones, with the number of dense, dark (i.e.  not containing stars),
substructures showing significant variation. In the second row, there are two
additional luminous satellites (see stellar image) close to the merging galaxies
that have no analogues in the other clones.

Such large variation in morphology between clones presents potential issues for
studies of the halo-galaxy connection in cosmological simulations of small
numbers of objects.  On the largest scales, there are relatively minor
differences between the haloes of these galaxies, with some substructure being
moved around. On the galactic scale, however, even if stellar masses are
similar, the CGM and specific merger timing of the galaxies remains
unconstrained.

We stress here that all of the galaxies predicted in different clones are
equally `accurate'. Although they all have slightly different feedback onset,
merger, and growth timescales, they are all valid predictions of the galaxy
formation model for such a history. These galaxies all live in extremely similar
large-scale environments, and in all cases constitute a $z=0$ system that is
either a pre-triple merger or an ongoing triple merger. As shown in Fig.
\ref{fig:scalingrelations}, the overall properties of galaxies are well
constrained between clones, further illustrated here with the similar stellar
masses. Tying the properties of the local CGM around such systems to the
properties of the environment or galaxy, however, requires investigation of many
haloes, or resimulation of the same galaxy (cloning, as here), to understand the
impact of stochastic variations on the results of the study.

Finally, we note that even though we attempted to match all 16 clones, only 14
matches were successful with our strategy. This was also the case with a traditional
particle matching method, and demonstrates the limitations of a one-to-one matching
strategy. As shown here, the most extreme cases are those that are the furthest
pre-merger, and in the unmatched two cases, the three major galaxies represented
here are entirely pre-merger, whereas in the dark matter-only case the halo is
entirely post-merger. To match this scenario, we would need to devise a strategy to
perform three-to-one matching, which was not anticipated before the onset of this work.
\section{Galaxy Property Scatter}
\label{sec:galaxypropertyscatter}

\begin{figure}
    \centering
    \includegraphics{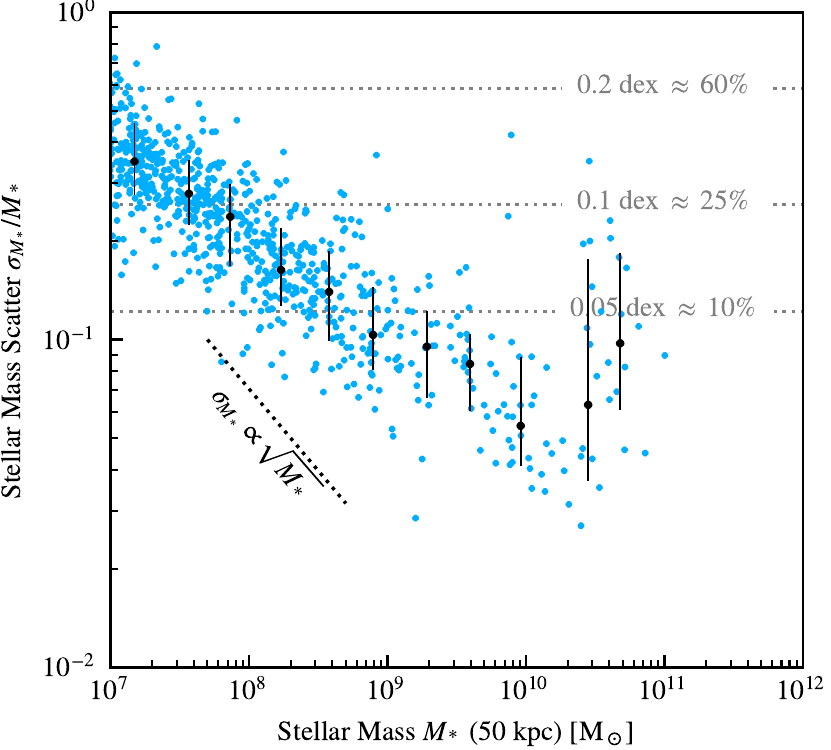}
    \caption{Scatter (using the same definition as Fig.
    \ref{fig:combinedspatialhalomassscatter}) in the galaxy stellar mass
    measured in a 3D 50 kpc aperture for all matched galaxies (blue points).
    Black points show median scatter in bins of $M_*$, with the error bars
    representing the 16-84 percentile range. The black dotted line shows the
    expected scaling (with the normalisation moved) if the scatter represented
    Poisson noise sampling for the stellar particles. The expected normalisation
    in this case would be approximately 0.1 dex at $M_* \approx 10^{7.5}$
    M$_\odot$ with the line moved for clarity.}
    \label{fig:allspatialstellarmassscatter}
\end{figure}

One of the most fundamental galaxy properties, used widely within carefully
designed scaling relations, is the galaxy stellar mass. Simulation suites are
now often explicitly tuned to reproduce the stellar mass of galaxies (either
through the stellar mass function alone, like EAGLE; \citealt{Crain2015}, or
additionally through the stellar mass-halo mass relation, like Illustris-TNG;
\citealt{Pillepich2018}). It is hence imperative that galaxy formation simulations
are able to reliably produce representative stellar masses, relative to their
input physics and parameters.

Fig. \ref{fig:allspatialstellarmassscatter} shows the scatter in galaxy stellar
masses, using 50 kpc 3D fixed stellar apertures  following
\citet{deGraaff2022}. The scatter in stellar mass is calculated as for halo
mass in Fig. \ref{fig:combinedspatialhalomassscatter}, representing half of the
16-84 percentile range across all clone simulations within which each
halo was matched.

In simulations of isolated galaxies, \citet{Keller2019} showed that the scatter
in stellar mass scaled as a Poisson-like error, with $\sigma_{M_*} \propto
\sqrt{M_*}$. For low-mass galaxies ($M_* < 10^{10}$ M$_\odot$), we find that
this is broadly true in the cosmological clones too, although the scatter scales
with stellar mass as approximately $\sigma_{M_*} \propto M_*^{-2/3}$. This
additional scatter in more realistic cosmological simulations is anticipated, as
\citet{Keller2019} found that scatter in stellar mass was significantly
increased during merger events which occur naturally in a cosmological
environment.

\begin{figure}
    \centering
    \includegraphics{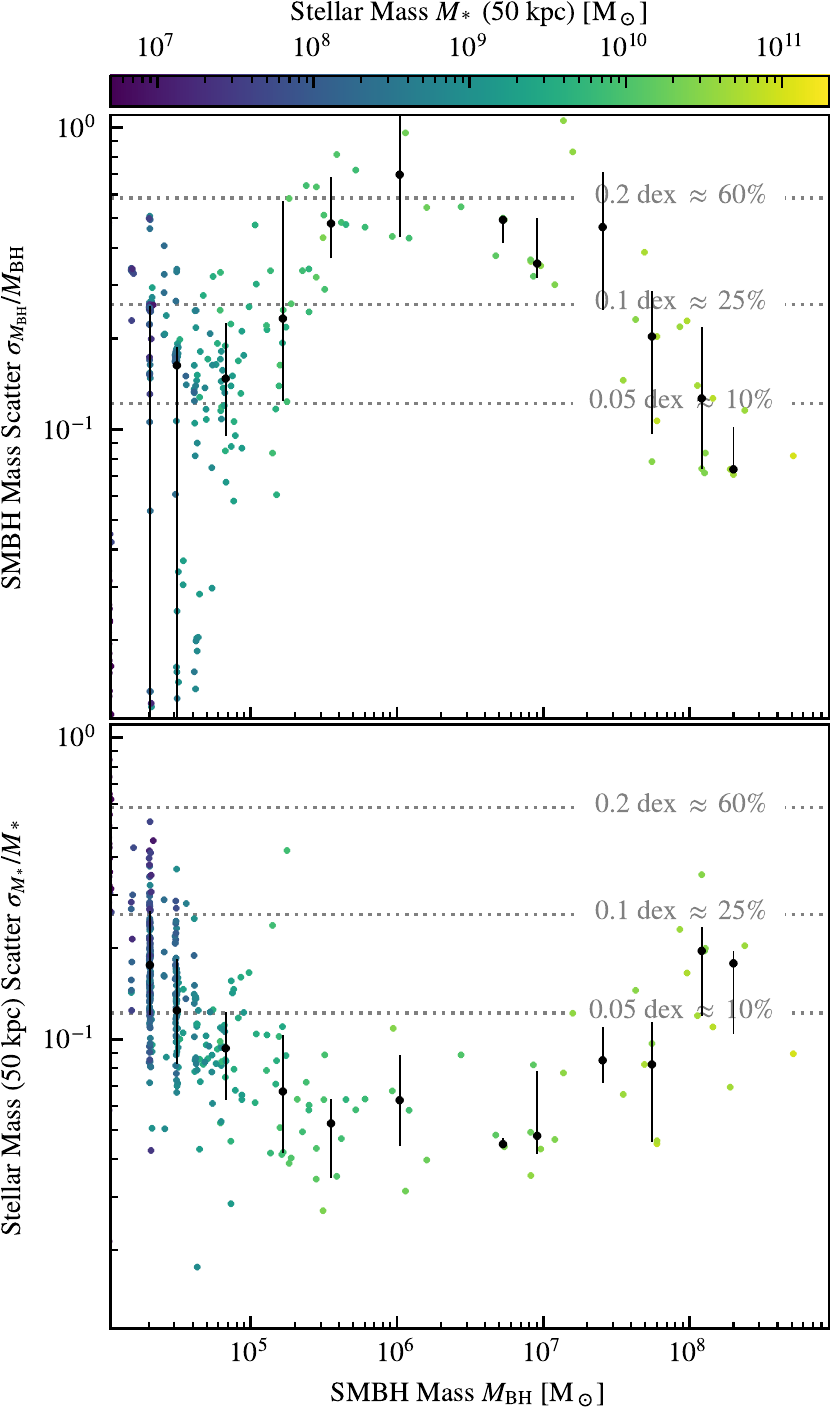}
    \caption{Scatter in SMBH mass (top) and stellar mass (bottom) as a
    function of SMBH mass. Points show individual matched galaxies, coloured by
    stellar mass, with black points and error bars showing the median and the
    16-84th percentile range. As black holes grow, their masses stabilise across
    clones simulations, but the associated highly energetic and stochastic AGN
    feedback leads to progressively larger scatter in the stellar mass of the
    host galaxies.}
    \label{fig:blackholescatter}
\end{figure}

The more surprising feature in Fig. \ref{fig:allspatialstellarmassscatter} is the
presence of an up-tick in scatter at galaxy masses of $M_* \gtrapprox 10^{10}$,
where feedback from active galactic nuclei (AGN) becomes significant
\citep{Bower2017, Bahe2022}.  To investigate the impact of black hole growth
(and its associated feedback) on galaxies, Fig. \ref{fig:blackholescatter} shows
the scatter associated with supermassive black holes (SMBHs). These are
identified as the most massive black hole resident in the host galaxy, and are
typically placed at the centre of the haloes, where they provide most of the AGN
feedback, through the recentering prescription \citep[see][]{Bahe2022}.

We first focus on the bottom panel, which shows the scatter in stellar mass (the
same as Fig. \ref{fig:allspatialstellarmassscatter}), but now as a function of
black hole sub-grid mass. At low black hole masses, we see relatively high
scatter in stellar masses due to these systems having a correspondingly low
stellar mass and hence high Poisson-like scatter (see above). As the black hole
mass grows beyond $M_{\rm BH} > 10^6$ M$_\odot$, the scatter in stellar mass
begins to level off and increase again at $M_{\rm BH} > 10^7$ M$_\odot$. 

In our model, black holes are seeded at $10^4$ M$_\odot$, and grow through
Bondi-Hoyle accretion of ambient gas. In the top panel of Fig.
\ref{fig:blackholescatter}, the scatter in black hole masses is shown, with high
levels of black hole scatter up to 100\% being sustained for black hole masses
$10^6 < M_{\rm BH} / {\rm M}_\odot < 10^8$ where these objects undergo rapid
growth relative to their host galaxy (see the stellar mass-black hole mass
relation in Fig. \ref{fig:scalingrelations}). As the onset of this rapid growth
has a random component (i.e. when the first dense gas settles within the
accretion radius), the timing of the growth phase is uncertain and hence quickly
leads to large variation in the black hole masses. In these cases, the black
hole mass is determined by relatively few accretion events, and as such each
accretion event comes along with significant black hole growth, making the
timing (which is stochastic) crucial. 

\citet{Genel2019} did not report an increase in stellar mass scatter at the
highest masses, although they found the same reduction in black hole mass
scatter at high masses that we report. This may be due to the different AGN
feedback implementations used in the two codes. For instance, the Illustris-TNG
model used by \citet{Genel2019} uses a kinetic radio mode, which may have a
different level of susceptibility to small timescale variations in gas
availability.

\begin{figure}
    \centering
    \includegraphics{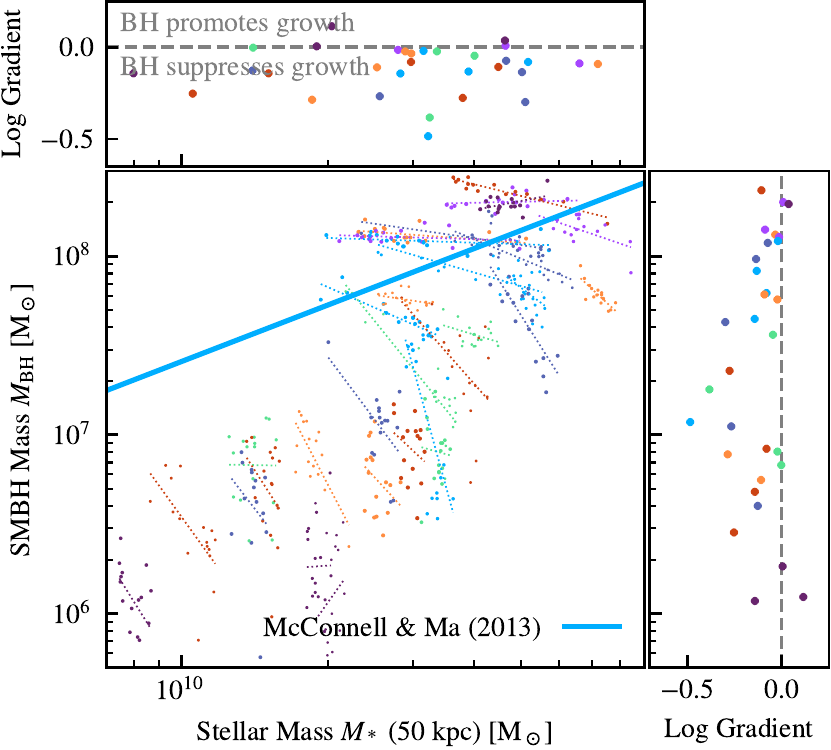}
    \caption{Zoomed-in region of the stellar mass-black hole mass relation (Fig.
    \ref{fig:scalingrelations}).  Individual colours now show single haloes
    matched between the clone simulations, with the dotted line showing a linear
    fit to the data (in log space). The panels to the top and right show, for
    each axis, the gradient of this fit as a function of stellar mass and black
    hole mass respectively. The grey dashed line in these two panels shows a
    gradient of zero, i.e. there is no dependence of the black hole mass on
    stellar mass (or vice-versa). Points lying above or right of the line
    indicate that the black hole mass grows as the stellar mass grows, whereas
    points that lie below the line indicate that the stellar mass
    in the halo is lower the higher the black hole mass is.}
    \label{fig:scatterindividalsmbhm}
\end{figure}

Fig. \ref{fig:scatterindividalsmbhm} shows the correlation between the scatter
between stellar mass and black hole mass between clones for individual haloes.
Each colour shows a single halo, in all clone simulations, and its scatter in
this space. To guide the eye, we fit straight lines to the scatter from each
halo, and show their gradients on the adjacent panels as a function of the two
variables. In the background we show the \citet{McConnell2013} data to show the
regime where galaxies are believed to self-regulate due to AGN feedback, with us
plotting their bulge mass data as a proxy of galaxy stellar mass meaning there
is an expected horizontal offset.

Across the mass range shown, black holes suppress the growth of galaxies.  There
is an anti-correlation between $M_{\rm BH}$ and $M_*$ within clones of
individual galaxies.  This leads to a negative gradient in $M_{\rm BH} / M_*$,
as indicated by the top and right panels of Fig. \ref{fig:blackholescatter}.

This figure begins to enlighten us on the trends that were observed in Fig.
\ref{fig:blackholescatter}, with galaxies at $M_* < 3\times10^{10}$ M$_\odot$
showing significantly larger scatter in black hole mass amongst clones than 
their higher mass counterparts, while more massive galaxies show instead an
increased diversity in their stellar mass across clones.

Between a black hole mass of $8\times10^6 < M_{\rm BH} / {\rm M}_\odot <
8\times10^7$, the gradient of $M_{\rm BH} / M_*$ indicates that the black holes
are strongly suppressing growth, as amongst clones lower black hole
masses correspond to more massive galaxies. In these scenarios, we would expect
the accretion history of the galaxies to be similar, but differences in initial
accretion (and hence feedback) timing lead to different output stellar masses
due to the changing onset times of quenching.

At the highest black hole masses, with $M_{\rm BH} > 10^8$ M$_\odot$, the
anti-correlation between black hole mass and galaxy mass weakens, with the
gradient of the $M_{\rm BH} / M_*$ relationship returning to $\approx 0$.

\begin{figure}
    \centering
    \includegraphics{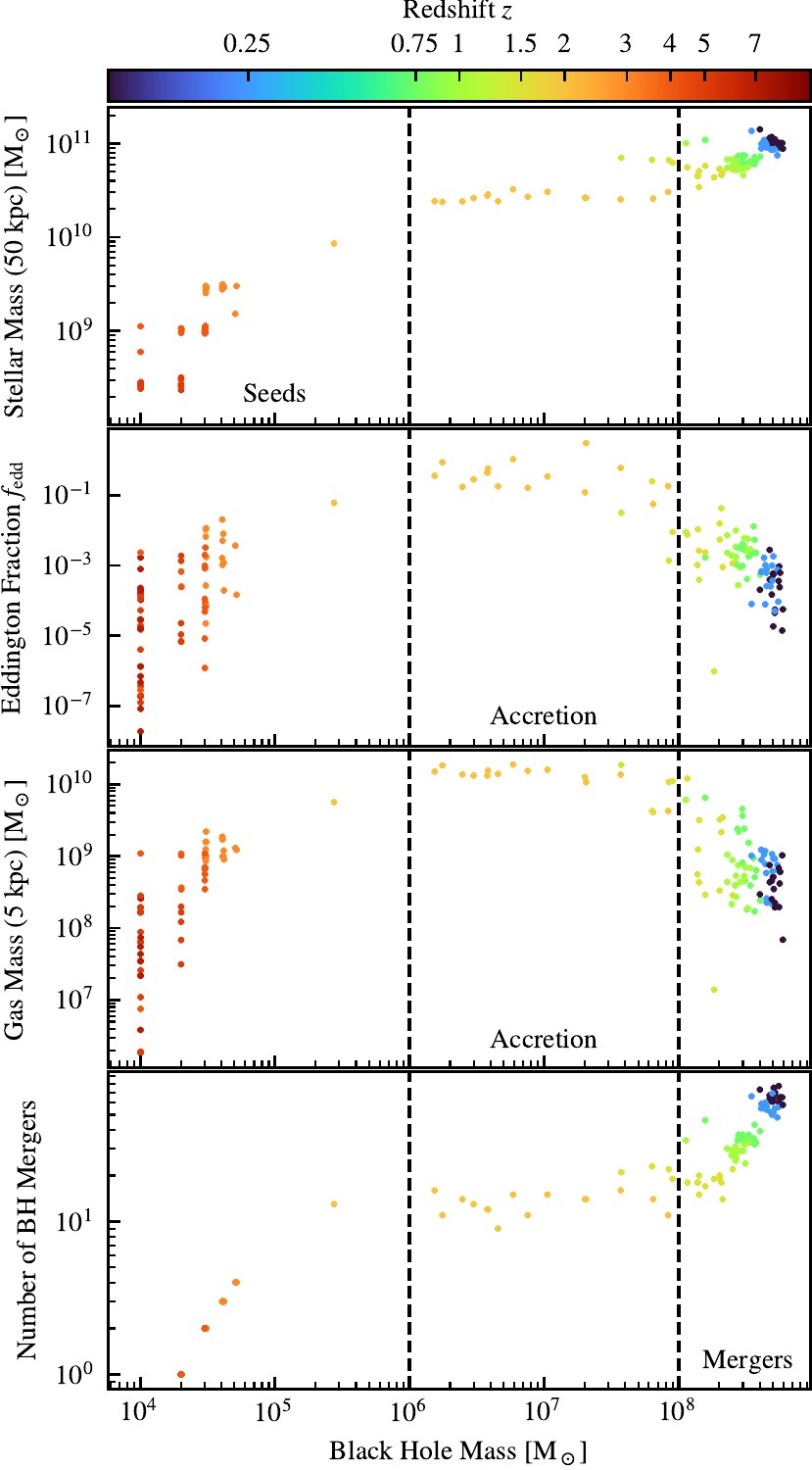}
    \caption{Tracks of a single clone galaxy (the most massive in the simulation volume),
    in four different key metrics. Each colour represents a different snapshot
    in time, with each point representing
    one clone of this galaxy. From top to bottom we show the stellar mass of the
    host galaxy, the Eddington fraction of the accreting black hole (at $f_{\rm
    Edd} = 1$, the black hole is maximally accreting), the gas mass within a
    fixed 5 kpc aperture around the black hole, and the total number of mergers
    experienced by the black hole. Two vertical dashed lines separate the three
    growth phases of the black hole, from it being a seed black hole, to it
    accreting close to its maximum rate, to finally merger-driven growth at
    masses $M_{\rm BH} > 10^8$ M$_\odot$.}
    \label{fig:blackholetrack}
\end{figure}

To investigate the relationship between the accretion onto black holes, the galaxy
they live in, and black hole growth, we now turn to tracking an individual
galaxy across both time and clones. In Fig. \ref{fig:blackholetrack}, we show
the history of the most massive galaxy in the volume, with a $z=0$ stellar mass
of $M_* \approx 10^{11}$ M$_\odot$. Each colour shows a single point in time,
with each point of a given colour showing the relationship between the property
and black hole mass in a given clone. We split the panels horizontally into
three segments, depending on the typical stage of black hole growth.

At masses $M_{\rm BH} < 10^6$ M$_\odot$, black holes are still effectively
seeds; they have not yet accreted the equivalent of one gas particle. The black
holes are accreting at significantly below the Eddington rate, with $f_{\rm Edd}
\lesssim 10^{-3}$, due to both the low black hole mass (Bondi accretion scales as
$\dot{M}_{\rm BH} \propto M_{\rm BH}^2$) and low surrounding gas density for
accretion, and indicating extremely long growth timescales through accretion.
Typically scatter in the black hole mass (i.e. horizontal scatter for a given
time) is driven by mergers with other seeds. Note how the scatter across black
hole mass is quantised at values of $M_{\rm BH} = \{1, 2, 3\} M_{\rm BH, seed} =
10^{4}$ M$_\odot$ (see also the number of mergers in the lower panel).

After roughly 10-20 mergers, the black holes have a high enough mass to accrete
efficiently from their surroundings. In addition, by this time, the gas density
has grown significantly, with in many cases there being $M_{\rm g} > 10^{10}$
M$_\odot$ within 5 kpc of the black hole \citep[see][]{Bower2017}. In this rapid
accretion phase ($f_{\rm Edd} \approx 1$), the black holes show significant
spread in mass (between $10^6 < M_{\rm BH} / {\rm M}_\odot < 10^8$), that is
roughly independent of galaxy stellar mass. At such high Eddington ratios, the
black holes can double their mass in less than 100 Myr, making the onset of such
rapid accretion a key determinant in their mass at a given point in time.

After the rapid accretion phase, gas expulsion from the galaxy core leads to
quenching of the galaxy, and shuts down black hole accretion (the rapid decline
in gas mass with black hole mass, green points). Once this happens, feedback
regulates the stellar masses of galaxies \citep[see][]{Booth2009}, with there
being a negative correlation between galaxy mass and black hole mass that
persists to $z=0$.  The increased scatter in stellar mass at the high mass end
of Fig. \ref{fig:allspatialstellarmassscatter} can therefore be understood in
terms of the stronger variation of AGN feedback. From this point onwards, the
Eddington fraction of the black holes is reduced to less than $f_{\rm Edd}
\approx 10^{-3}$, with black holes undergoing many more mergers, indicating that
they grow much more slowly. As shown in \citet{Bahe2022}, these mergers are
mainly with seed-mass black holes, and do not significantly contribute to the
mass of the central SMBH.

This case study also highlights how studies on the stochasticity of the model
can help us gain insight into the inner workings of the model due to the small
changes in accretion history and merger timings they create.
\section{Conclusions}
\label{sec:conclusions}

Cosmological galaxy formation simulations employ a suite of stochastic sub-grid
models to coarse-grain continuous physical processes. These stochastic models
enable physics that usually acts on small timescales, and on small masses
(relative to the time-step and particle mass resolution in the simulation) to be
approximated in an effort to make simulations computationally feasible. In this
paper, we have closely examined the impact of these stochastic models on both
the resultant global predictions and properties of individual galaxies simulated
with the SWIFT code and SWIFT-EAGLE model through 16 re-simulations of the same
$25^3$ Mpc$^3$ volume labeled as `clones'. Our main findings are as follows:
\begin{itemize}
    \item In Fig. \ref{fig:scalefactorupdates}, we showed that modern,
          non-deterministic, task-based codes inevitably lead to deviations between
          simulations after roughly 1000 steps, or 300 Myr into the simulation,
          for a 25 Mpc volume. All of these re-simulations are equally valid (or
          invalid): these deviations arise from round-off errors that accumulate
          in both the gravity and hydrodynamics calculations, leading to differences
          in the timings of the first stars and hence first feedback events.
    \item In Figs. \ref{fig:globalpropertiessfr} and
          \ref{fig:globalpropertiesbhard}, we showed how even globally averaged
          properties in a full cosmological volume differ between clone
          simulations. At a fixed redshift, the star formation rate density is
          only constrained to $\approx 0.1$ dex, and the black hole accretion
          rate density only to within $\approx 1.0$ dex, in a 25 Mpc volume. In
          larger simulation volumes these differences in globally averaged
          quantities between clones are expected to diminish even further.
    \item In Fig. \ref{fig:num_substructure}, we showed that the number
          of substructures within individually matched FoF haloes is well
          constrained between clone simulations, especially at masses $M_{\rm
          200, crit} > 10^{11}$ M$_\odot$, with the scatter between clone
          galaxies reducing as mass increases. This metric is likely to be
          significantly affected by the resolution of our simulations, but
          it confirms that the accretion history for the most massive haloes is
          well constrained between clone simulations.
    \item In Fig. \ref{fig:scalingrelations}, we investigated the impact of
          differences between clone simulations on our ability to predict scatter in
          scaling relations with simulations. In some cases, the scatter in
          scaling relations can be significantly affected by the variability in
          the properties of individual galaxies.  This typically occurs at the
          lowest galaxy masses, with the clone-to-clone scatter dominating at at
          $M_* \lessapprox 10^{8.5}$ M$_\odot$, but high levels of variability can
          occur at high masses where galaxies are well resolved when the
          property is time-sensitive (e.g. sSFR and galaxy quenching).
    \item In Figs. \ref{fig:differentialmassgrowth} and \ref{fig:casestudy},
          we presented a case study of one individual galaxy across the clone
          simulations with $M_* \approx 2\times10^{10}$ M$_\odot$ to investigate
          the origin of scatter in its properties. Although global properties,
          such as halo and stellar mass, are well constrained (showing around
          10\% variation), detailed properties such as the positions of
          substructure, phase structure of the ISM, CGM, and even IGM were
          different between clones. Investigations into the
          halo-galaxy connection, particularly those employing such phase
          information, must use adequate sampling to suppress the random
          variability within the galaxy formation model.
    \item In Fig. \ref{fig:scatterindividalsmbhm} we investigated the
          correlation between scatter in galaxy stellar mass and black hole mass
          to identify the cause of an increase in clone-to-clone stellar mass
          scatter at $M_* \approx 3 \times 10^{10}$ M$_\odot$. There is a strong
          anti-correlation between black hole and stellar mass of individual
          galaxies across clone simulations in the mass range $10^6 < M_{\rm BH}
          / {\rm M}_\odot < 10^8$, which weakens at higher black hole masses.
\end{itemize}
In summary, we have found that whilst the properties of individual galaxies can
be significantly impacted by random variations between clones, and hence the
stochasticity of modern galaxy formation models, large-scale properties such as
galaxy scaling relations are relatively robust against this random variability.
The medians of these scaling relations are well constrained, but there is a
significant component of the scatter, especially in the poorly-resolved
(low-mass) regime, that originates from differences due to random variability.
At higher masses, galaxy mergers and those with recent strong feedback events
can see huge variation in their measured properties between clone simulations.

It is likely that the differences between clones are driven by the choice to
discretise sub-grid models stochastically. Although there may be some component
of the scatter that would still remain if all models were continuous (mainly
driven by round-off errors causing different timing for e.g. AGN feedback), the
residual differences would likely be small, as shown in the appendix of
\citet{Genel2019}. Despite the additional variability created in the final
galaxy properties by these models, they remain extremely useful due to their
small memory footprint and their immunity to spurious gravitational heating
effects \citep{Ludlow2021}.

These results indicate that analysts of individual objects must be cautious
about the unpredictability of galaxy properties due to random variations,
notably in certain setups when model parameters are changed, or in constrained
realisation simulations aiming to reproduce properties of a given observable
object.

Future work should investigate the impact of random variability on satellite
galaxies, as well as reconsider the one-to-one matching paradigm that we have
explored here; in many cases, the most impacted objects are those that are split
into many (generally pre-merger) within clones. It should also consider
additional sources of stochasticity (those not based upon random numbers) that
were unexplored here, such as stochastic gravitational scattering and the
influence of high mass ratios between baryonic and dark matter resolution
elements.

\section{Acknowledgements}

JB and the authors gratefully acknowledge the significant contribution
to this paper that Richard G. Bower made. He provided mentorship and
advice on this and many other related works, but was unable to take part
in the final stages of preparation for this work due to illness. This paper
would not have been possible without his expert guidance and support.

The authors thank the anonymous referee for their helpful comments.
The authors acknowledge helpful conversations with Ben Keller.
YMB gratefully acknowledges funding from the Netherlands Organization for
Scientific Research (NWO) through Veni grant number 639.041.751  and financial
support from the Swiss National Science Foundation (SNSF) under funding
reference 200021\_213076.
ADL acknowledges financial support from the Australian Research Council through
their Future Fellowship scheme (project number FT160100250).
EA acknowledges the STFC studentship grant ST/T506291/1.
This work used the DiRAC@Durham facility managed by the Institute for
Computational Cosmology on behalf of the STFC DiRAC HPC Facility
(www.dirac.ac.uk). The equipment was funded by BEIS capital funding
via STFC capital grants ST/K00042X/1, ST/P002293/1, ST/R002371/1 and
ST/S002502/1, Durham University and STFC operations grant
ST/R000832/1. DiRAC is part of the National e-Infrastructure.

\subsection{Software Citations}

This paper made use of the following software packages:
\begin{itemize}
    \item {\sc Swift}: \citep{Schaller2018}
    \item {\sc VELOCIraptor}: \citep{Elahi2019}
	\item {\textsc{Matplotlib}}: \citet{Hunter2007}
	\item {\textsc{SciPy}}: \citet{Virtanen2020}
	\item {\textsc{NumPy}}: \citet{Harris2020}
	\item {\textsc{unyt}}: \citet{Goldbaum2018}
	\item {\textsc{SwiftSimIO}}: \citet{Borrow2020,Borrow2021a}
\end{itemize}

\section*{Data Availability}

All data generated and used in this article were generated using the open source
simulation code SWIFT\footnote{Available at \url{swiftsim.com}.}, and use the
publicly available VELOCIraptor code\footnote{Available at
\url{https://github.com/ICRAR/VELOCIraptor-STF}} for structure finding. The
SWIFT-EAGLE model is freely available in the SWIFT repository, as are the
initial conditions. All data analysis and reductions were performed with
swiftsimio and related open source tools\footnote{Available at
https://github.com/swiftsim}. The data volume of the simulations is large ($\sim
3$ TB), but the underlying computing time to perform them is modest (less than
100'000 CPU hours).
 


\bibliographystyle{mnras}
\bibliography{example} 








\bsp	
\label{lastpage}

\end{document}